\documentclass[useAMS,usenatbib]{mn2e}

\def\equ#1{eq.~(\ref{eq:#1})}
\def\Equ#1{Eq.~(\ref{eq:#1})}
\def\se#1{\S\ref{sec:#1}}
\def\Fig#1{Fig.~\ref{fig:#1}}
\def\etal{{\it et al.\ }}

\def\la{\langle}

\def\be{\begin{equation}}
\def\ee{\end{equation}}
\def\prop{\propto}
\def\ifm#1{\relax\ifmmode#1\else$\mathsurround=0pt #1$\fi}
\def\kms{\ifmmode\,{\rm km}\,{\rm s}^{-1}\else km$\,$s$^{-1}$\fi}
\def\hmpc{\,\ifm{h^{-1}}{\rm Mpc}}

\def\kpc{\,{\rm kpc}}
\def\d{{\rm d}}

\def\msun{M_{\odot}}

\def\ltsima{$\; \buildrel < \over \sim \;$}
\def\lsim{\lower.5ex\hbox{\ltsima}}
\def\gtsima{$\; \buildrel > \over \sim \;$}
\def\gsim{\lower.5ex\hbox{\gtsima}}
\def\const{{\rm const.}}

\def\omm{\Omega_{\rm m}}

\def\Vmax{V_{\rm max}}

\def\Mg{M_{\rm g}}
\def\Ms{M_{\rm *}}
\def\Ls{L_{\rm *}}
\def\fb{f_{\rm b}}
\def\fd{f_{\rm d}}

\def\Rs{R_{*}}

\def\sfr{\dot{M}_*}
\def\trad{t_{\rm rad}}
\def\tff{t_{\rm ff}}
\def\Esn{E_{\rm SN}}
\def\Vsn{V_{\rm SN}}
\def\Msn{M_{\rm SN}}
\def\Mssn{M_{*{\rm SN}}}

\def\mus{\mu_*}
\def\mug{\mu_{\rm gas}}

\def\zion{z_{\rm ion}}
\def\Tion{T_{\rm ion}}
\def\Vev{V_{\rm evap}}
\def\Vjeans{V_{\rm Jeans}}
\def\fbound{f_{\rm bound}}

\def\pmb#1{\setbox0=\hbox{#1}%
\kern-.025em\copy0\kern-\wd0
\kern.05em\copy0\kern-\wd0
\kern-.025em\raise.0433em\box0}

\title[Feedback in dwarf/LSB galaxies]
{Feedback and the fudamental line of low-luminosity LSB/dwarf galaxies}

\author[A. Dekel \& J. Woo]
{Avishai Dekel$^{1}$ \& Joanna Woo$^{1,2}$\\
$^1$Racah Institute of Physics, The Hebrew University, Jerusalem
91904, Israel\\
$^2$Department of Physics \& Astronomy, University of British Columbia, 
BC V6T 1Z1, Canada}

\begin{document}

\pagerange{\pageref{firstpage}--\pageref{lastpage}} \pubyear{2003}
\maketitle
\label{firstpage}

\begin{abstract}
We study in simple terms the role of feedback in establishing the scaling 
relations of low-surface-brightness (LSB) and dwarf galaxies 
with stellar masses in the range 
$6\times 10^5 \leq\Ms \leq3\times10^{10}\msun$.
These galaxies, as measured for example from SDSS and in the Local Group,
show tight correlations of 
internal velocity, metallicity and surface brightness (or radius) with $\Ms$. 
They define a {\it fundamental line} which distinguishes them from the 
brighter galaxies of high surface brightness and metallicity. The idealized 
model assumes spherical collapse of CDM haloes to virial equilibrium and 
angular-momentum conservation.  The relations for bright galaxies are 
reproduced by assuming that $\Ms$ is a constant fraction of the halo mass $M$. 
The upper bound to 
the low-luminosity 
LSBs coincides with the virial velocity of haloes in which 
supernova feedback could significantly suppress star formation, $V<100\kms$ 
(Dekel \& Silk 1986). We argue that the energy fed to the gas obeys 
$\Esn\prop\Ms$ despite the radiative losses, and equate it with the binding 
energy of the gas to obtain $\Ms/M\prop V^2$. This idealized model provides 
surprisingly good fits to the scaling relations of 
low-luminosity LSBs and dwarfs, 
which indicates 
that supernova feedback had a primary role in determining the fundamental line.
The apparent lower bound for galaxies at $V \sim 10\kms$ may be due to the 
cooling barrier at $T\sim10^4$K.  Some fraction of the dark haloes may show no 
stars due to complete gas removal either by supernova winds from neighboring 
galaxies or by radiative feedback after cosmological reionization at $\zion$.
Radiative feedback may also explain the distinction between dwarf spheroidals 
(dE) and irregulars (dI), where the dEs, typically of $V\leq30\kms$, form stars
before $\zion$ and are then cleaned out of gas, while the dIs, with $V>30\kms$,
retain gas-rich discs with feedback-regulated star formation.
\end{abstract}


\section{Introduction}
\label{sec:intro}

The galaxies can be crudely divided into two main classes based on their 
location in the plane of surface brightness versus luminosity 
or stellar mass $\Ms$
[see for example Fig.~1 in Dekel \& Silk (1986, hereafter DS), 
Fig.~7a of Kauffmann \etal (2003b, hereafter K03), and other   
references cited below].        
The bright galaxies, dominated at the bright end by ellipticals 
and early-type spirals, 
have relatively high surface brightnesses which are only weakly
correlated with stellar mass. The fainter galaxies, 
spanning the broad range from relatively bright late-type spirals 
all the way down to the Local-Group dwarf galaxies, have their 
conditional 
average surface brightness at a given $\Ms$ decrease with decreasing $\Ms$. 
The {\it transition} occurs near $\Ms \simeq 3\times10^{10}\msun$ 
(corresponding to absolute magnitude of about $-20.8$ and $-19.0$
in the r and b bands respectively).
Other global galaxy properties, such as the mean metallicity, behave in a 
similar manner as a function of stellar mass, with the transition seen at 
a similar characteristic scale 
(see references below). 
We refer to these two general classes hereafter as HH versus LL 
galaxies, standing for High-luminosity High-surface-brightness (HSB) 
galaxies versus Low-luminosity Low-surface-brightness (LSB) and dwarf 
galaxies. 
For the purpose of our current idealized theoretical modeling,
we simply distinguish between these two coarse types based on the 
$\Ms$ transition scale.\footnote{LSBs are sometimes defined in the 
literature by central blue surface brightness $>22$ mag arcsec$^{-2}$, 
1-$\sigma$ from Freeman's (1970) mean value for bright HSB spirals 
($21.65 \pm 0.35$), 
and sometimes by $>23$ mag arcsec$^{-2}$. 
By referring to LL and HH we do not imply that there are no LSB 
galaxies of high luminosity (see below). 
The term ``dwarfs" is also used in different ways, 
referring alternatively to small, faint, and low surface brightness galaxies.
}
This kind of classification can be traced back, e.g., to  
Binggeli, Sandage \& Tarenghi (1984), 
Wirth \& Gallagher (1984), Kormendy (1985) and Hoffman \etal (1985).
Dekel \& Silk (1986) have highlighted this classification
scheme (their Fig.~1) in the context of their early
theoretical modeling (their Fig.~6).

\subsection{HH and LL galaxies} 

The analysis by K03 of 80,000 galaxies 
from the Sloan Digital Sky Survey (SDSS) highlight the
bivariate distribution of relatively bright galaxies in the plane 
of surface brightness and stellar mass, 
above their claimed completeness limits of absolute r magnitude $-17$ 
and effective surface brightness $23$ mag arcsec$^{-2}$.
With the spectral information available for SDSS galaxies, their stellar
masses can be evaluated more reliably than before
using population synthesis models (Kauffmann \etal 2003a).
The transition scale shows very clearly at 
$\Ms \simeq 3 \times 10^{10} \msun$ (K03, Figure 7a).
The bright galaxies in the range $3\times 10^{10} < \Ms < 10^{12} \msun$
have their effective surface brightnesses scattered 
about a mean value of $\mus \sim 10^{9} \msun \kpc^{-2}$
(referring to the mean surface brightness within the half-light radius; 
the central surface brightness is typically larger by a factor of $\sim 3$),
with only a weak systematic trend of roughly $\mus \prop \Ms^{0.2}$. 
(When viewed as a function of luminosity, the surface brightness is actually
decreasing slowly with luminosity in this range, because the stellar
mass-to-light ratio is increasing, see Blanton \etal 2003.)
On the other hand, 
the correlation at the top part of the 
LL 
regime, $10^8 < \Ms < 3 \times 10^{10} \msun$,
is well fit by $\mus \prop \Ms^{0.6}$ (or even slightly steeper).
A similar correlation, with a slope $\simeq 0.6-0.7$, is found in the top 
LL 
regime between surface brightness and $i$ absolute magnitude based on
144,609 SDSS galaxies (Blanton \etal 2003), indicating that the translation 
to stellar mass by K03 makes a negligible difference in this regime.
A consistent correlation is measured from other samples of galaxies as well,
e.g., Cross \etal (2001) find a slope of $\simeq 0.42$ in the 2dF survey
over the whole range of galaxies brighter than $M_{\rm b} \simeq -16$,
de Jong \& Lacey (2000) measure a slope of $0.5$ for Sdm galaxies,
Driver (1999) finds a slope of $0.67$ for a sample of fainter galaxies
in the Hubble Deep Field,
and Ferguson \& Binggeli (1994) reported a slope of $0.7$ for Virgo dwarf
galaxies.

The spread in surface brightness at a given luminosity, which is not
directly relevant for our theoretical analysis in the current paper,
is a matter of debate among the observers.
The uniformly selected SDSS data show a relatively tight distribution 
about the mean relation in the $\mus-\Ms$ plane, both above
and below the transition scale (see K03, Figs.~7, and 10).
We learn for example that low-luminosity galaxies with high 
surface brightness, such as M32, seem to be rare. 
If the photometric completeness limit of the SDSS data in r
is indeed below $23.0$ mag arcsec$^{-2}$, 
then, for a given $\Ms$ near $\Ms \sim 10^9\msun$, 
this data indicate a significant drop in 
the galaxy count as a function of decreasing surface 
brightness.\footnote{Blanton 
\etal (2003, in preparation) provide further evidence for
their completeness down to below $23.0$ 
by showing that when the galaxies are binned
according to Galactic extinction $E$, the peak of the apparent
surface brightness distribution shifts accordingly,
from about 21.2 at $E=0$ to about 23.0 at $E=1.8$
(a test first applied by Davies \etal 1993 to the ESO-Uppsala catalog).
} 
A similar conclusion, of a dearth of high-luminosity galaxies with
low surface brightness, is obtained from the 2dF survey by Cross \etal (2001).
On the other hand, there are claims in the literature for a significant 
population of low surface brightness galaxies with high luminosities
(e.g., Disney 1976; Phillipps, Davies \& Disney 1988, 1990; Davies \etal 1994;
Impey \etal 1996; Bothun, Impey \& McGaugh 1997; Sprayberry \etal 1997;
papers in Davies, Impey \& Phillipps 1999;
O'Neil \& Bothun 2000).\footnote{
O'Neil \& Bothun 
argue for a population of 
luminous LSBs based on their finding that the surface brightness distribution 
function is flat down to below a central blue value of $24$ 
-- down to the survey completeness limit. 
They infer that many of the LSBs are luminous and extended
based on the moderate differences in the distributions of sizes and 
velocities between HSBs and LSBs in these surveys. 
With the correlation $\mus \prop \Ms^{0.6-0.7}$, the expected variation in 
radius across this decade of $\mus$ is only a factor of $\sim 2$, 
which may be consistent with the data. 
The corresponding variation in velocity, $V \prop \Ms^{0.2}$,
is similarly weak. The flat surface-brightness distribution can therefore
be consistent with the correlation detected in SDSS, 2dF and other surveys
between surface brightness and luminosity.
}
The SDSS and 2dF results argue that this is either a smaller population,
or a separate population that occupies a different
locus in the $\mus-\Ms$ plane, below the surface-brightness limits of
these surveys. 
Independently of this ongoing dispute, we seek in
the current paper theoretical understanding for the general correlation 
between average surface brightness and stellar mass,
as indicated by SDSS and the other datasets across 5 decades in $\Ms$,
from the transition scale to the smallest dwarfs.
As mentioned in \se{conc},
the scatter in surface brightness at a given luminosity may be affected
by other physical processes not studied here in detail --- the obvious one 
being angular momentum (e.g., Dalcanton, Spergel \& Summers 1997).

We notice in passing that, given the observed Schechter luminosity function 
and the transition at $\Ms \simeq 3 \times 10^{10} \msun$ (about a factor
or 2 below the mass corresponding to Schechter's characteristic $\Ls$),
more than 95\% of the galaxies are below the transition scale,
while most of the light still comes 
from the bright galaxies above this scale.  
In terms of mass, if the mass function of haloes is similar 
to that predicted for the $\Lambda$CDM cosmology by simulations or
Press-Schechter-like approximations, then the vast majority of 
the virialized mass is in haloes of 
LL 
galaxies.  This clearly motivates a major theoretical effort aimed at 
understanding the origin of the 
low-luminosity 
LSBs.

At least two additional independent relations, beyond the $\mus-\Ms$ relation,
are apparent in the SDSS data 
of the relatively bright galaxies. 
The second, based on preliminary reports (Tremonti \etal 2003, in preparation),
is a re-confirmation of a scaling relation involving the metallicity $Z$,
of roughly $Z \prop \Ms^{0.5}$ at the high end of the 
LL 
regime, turning into no significant correlation between $Z$ and $\Ms$ in the 
HH 
regime, $Z\simeq$const.
Similar gradients in gas metallicity have been seen before in other samples 
of LSBs (e.g., McGaugh 1994; de Blok \& McGaugh 1997). 

The third, which is typically the tightest correlation obeyed by galaxies, 
is between their luminosity and the characteristic velocity $V$ 
measuring the depth of the gravitational potential well, corresponding to
roughly $V \prop \Ms^{1/4}$ (e.g., Bernardi \etal 2003).
This is the Tully-Fisher relation for the rotation velocity in discs 
and the Faber-Jackson relation (or a projection of the generalized 
Fundamental-Plane relation)
for the dispersion velocity in spheroids. We term this kind of relation
between $\Ms$ and $V$ a ``TF" relation.  
A similar TF relation seems to extend down at least to the top part of 
the 
LL 
regime, with no obvious change at the transition between 
HH and LL 
galaxies (e.g., Zwaan \etal 1995; Sprayberry \etal 1995; Dale \etal 1999).

K03 (Fig.~9 and 11) also find that the luminosity 
``concentration" within the galaxies correlates with stellar mass,
which they interpret as a measure of bulge-to-disc ratio.
The bright galaxies are dominated by elliptical galaxies, the 
LLs 
near $\Ms \sim 10^{9}\msun$ are dominated
by discs, and in between the bulge-to-disc ratio is gradually decreasing with 
decreasing $\Ms$. Associated with this trend is an increasing gas-to-star
ratio (e.g., Longmore \etal 1982; McGaugh \& de Blok 1997)  
and a younger, bluer stellar population in galaxies of decreasing $\Ms$
down to $\sim 10^8\msun$ 
(e.g., de Blok, van der Hulst \& Bothun 1995; 
Sprayberry \etal 1995; McGaugh, Schombert \& Bothun 1995; 
another interpretation in van den Hoek \etal 2000).
At a given $\Ms$, the galaxies with lower bulge-to-disc ratio and younger
stellar ages tend to be of lower surface brightness 
(K03, Fig.~14).
Preliminary results from the SDSS data (Brinchmann \etal 2003, in preparation)
indicate that while
the current star formation rate in HHs shows no clear correlation with
stellar mass, there is a correlation of the sort $\sfr \prop \Ms$ in the 
top 
LL 
regime.

\subsection{Local-Group dwarfs}

The {\it dwarf galaxies} of the Local Group (LG, see \se{LSB_LG}), 
as well as those observed in the Virgo cluster and the Local Supercluster
(e.g., Binggeli \& Cameron 1991; Ferguson \& Binggeli 1994), 
seem to be in a crude sense an extension of the 
LL 
population of SDSS, 
stretching its range to almost 5 decades in stellar mass:
$6\times 10^5 < \Ms < 3\times 10^{10}\msun$. The LG dwarfs themselves span
almost the whole 
LL 
range, up to $\sim 10^{10}\msun$,
and they therefore constitute a very useful sample,
which we use extensively in this paper.
As summarized below and displayed in figures in \se{LSB_LG},
the correlations between the observed global quantities seem to be roughly
consistent throughout this range, including the surface brightness, metallicity
and velocity as a function of stellar mass. 
The correlations between every pair of parameters for galaxies in this regime
define a {\it fundamental line} in the multi-parameter space, say $\Ms$,
$\mus$, $V$ and $Z$.
For the purpose of our idealized modeling of this mean line,
we ignore the differences between different types within the 
galaxies that populate the line, 
and any possible population outside this family.
%
 
Within the broad 
LL 
family,
the brighter dwarfs are typically disc-like or irregular in shape and gas rich
(we use hereafter the broad term ``dwarf irregulars", dI);  
in many ways they seem to represent 
a continuous extension of the 
LLs 
observed by SDSS.
On the other hand, the faint end, $\Ms \leq 3\times 10^7\msun$,
is dominated by dwarf ellipticals or
spheroidals (hereafter dE), typically with only little gas and current 
star formation.
While the dEs extend the general 
LL 
trends of decreasing
surface brightness and metallicity, they clearly represent a reversal in
the trend of
galaxy type (bulge-to-disc ratio, gas-to-stars ratio and star-formation rate),
and as we show below
they seem to show an additional difference in the velocity trend (TF).  
After addressing the origin of the 
LL/dwarf 
family as a whole in the main body of
this paper, we also discuss the possible origin of the distinction between dEs 
and dIs.

The data on the $\sim 40$ dwarf galaxies of the Local Group, 
based primarily on the compilations by Mateo (1998)
and van den Bergh (2000), have been analyzed by 
Woo \& Dekel 
(in preparation, hereafter WD). 
They computed the corresponding 
scaling relations in the range $6\times 10^5 < \Ms < 10^{10}\msun$ 
in comparison with the SDSS results for the bright end of the 
LLs. 
The stellar mass $\Ms$ of each galaxy has been derived from the observed 
magnitudes using the mean age and metallicity of the stellar population
and a simple population synthesis model (kindly provided by G. Kauffmann);
%
the results were found to be quite insensitive to the details of this 
derivation.
For the central surface brightness 
WD find a tight correlation about the scaling relation
$\mus \prop \Ms^{0.55 \pm 0.03}$, 
extending down to $\mus \sim 3\times 10^6\msun \kpc^{-2}$ at the faint end.
The best-fit slope is determined by WD via a linear regression of the log 
variables,
taking into account the errors in both (i.e., minimizing the 2-dimensional
$\chi^2$ as in Numerical Recipes, Press \etal 1992, \S 15.3).
The Pearson correlation coefficient (NR, eq.~14.5.1) is $r=0.88$.
The slope is quite similar to the slope obtained for the effective surface
brightness of 
LLs 
sampled by SDSS, while the amplitude for the effective
$\mus$ is lower by a factor of $\sim 3$.

For the metallicity $Z$, WD took the stellar
[Fe/H] (mostly for dE) and/or a constant factor times the Oxygen abundance
of the gas
(mostly for dI), where the constant factor has been chosen to minimize
the scatter in the $Z-\Ms$ relation. They find a tight correlation,
with the best-fit scaling relation $Z \prop \Ms^{0.40\pm 0.02}$ and $r=0.92$
 
For the internal velocity $V$, WD adopted
the observed maximum circular velocity for the dIs 
and $\sqrt{3}\sigma_{\rm p}$ for dEs, 
where $\sigma_{\rm p}$ is the observed projected central dispersion velocity.
When the fit is performed across the whole dwarf range, 
the TF scaling relation is $V \prop \Ms^{0.24 \pm 0.01}$ with $r=0.89$.
When inspected more carefully, the dIs at the bright end show a slight
steepening which merges smoothly into the known TF relation for bright 
galaxies.
At the faint end, $\Ms < 3\times 10^7\msun$,
there is an indication that the velocities of the dEs are bound from below
by $V \geq 10\kms$, and can actually be fit by $V \simeq {\rm const.}$
(see \se{rad}). Nevertheless,
the tight scaling relations over the whole range indicate that the 
LL/dwarf 
galaxies basically constitute a one-parameter family, which calls for a 
simple physical explanation.

\subsection{Supernova feedback}

The dwarf galaxies are central players in one of the main problems
facing galaxy-formation theory in the context of CDM cosmology --- 
the so called ``missing dwarf problem".
This refers to the apparent discrepancy between the predicted abundances
of halo masses in the CDM cosmology, especially subhaloes within larger haloes,
and the relatively few, faint dwarf galaxies observed, e.g.,  
in the Local Group (Klypin \etal 1999; Moore \etal 1999; 
Springel \etal 2001). 
This problem is related to the fact that
for galaxies fainter than $\Ls$ the luminosity function is observed
to be flatter than the mass function predicted for haloes in the 
$\Lambda$CDM cosmology. 
This implies that the stellar-to-virial mass ratio $\Ms/M$ must be 
decreasing with decreasing $M$, namely fewer stars were formed per unit 
total mass in fainter LSB galaxies. 
The systematic variation in $\Ms/M$ can serve as a clue for 
understanding the 
LL-LSB 
phenomenon.
Following preliminary ideas by Larson (1974), 
Dekel \& Silk (1986) studied the general
scenario where the key physical process governing this phenomenon
is the {\it supernova feedback} from a first generation of stars, which 
either drives out a significant fraction of the original halo gas 
or suppresses star formation in the retained and added gas.
DS showed that the observed scaling properties of dwarf galaxies
{\it can} be qualitatively consistent with
this picture, provided that the potential wells are dominated by non-gaseous
dark haloes with a structure that crudely resembles the predictions
of the CDM scenario. They
studied the amount of energy fed into the interstellar gas by supernova
ejecta subject to radiative losses and found that haloes with virial velocities 
significantly lower than a critical value of order $\sim 100\kms$ can lose 
significant amounts of their gas and/or effectively suppress further star 
formation.

Recent observations provide cumulative evidence for massive
outflows from galaxies, consistent with being generated by supernovae. 
For example, HI maps of large disk galaxies show empty bubbles associated
with outflowing clouds, which indicate an energy equivalent to $\sim 100$ 
normal supernovae (Boomsma \etal 2002).
Chandra x-ray measurements show extended winds of soft x-ray about galaxies
such as M82, associated with hard x-ray sources in the galaxy, indicative of
young massive stars and supernovae (Roy \etal 2000; Martin, Kobulnicky \&
Heckman 2002). 
Outflows are detected directly in local starburst galaxies,
and 
are 
seen to be driven by SNII activity (e.g., Legrand \etal 1997; 
Martin 1999b; Heckman \etal 2001).  
In some of these galaxies, outflows of a few hundred $\kms$ are inferred 
based on the blueshifted metal absorption lines from the approaching 
foreground compared to the redshifted Ly-$\alpha$ photons backscattered
from the receding background.
Spectroscopy of the brighter lensed galaxies at high redshifts reveal
using a similar effect typical outflows of $200-800\kms$
(Franx \etal 1997; Frye \& Broadhurst 1998; 
Pettini \etal 2001; Frye, Broadhurst \& Benitez 2002).
New evidence for strong outflows at $z\sim 3$ is provided by 
Adelberger \etal (2003), who interpret their measurements of absorption
systems in QSO spectra as bubbles of radius $\sim 1\hmpc$
(comoving) around Lyman-break galaxies, almost empty of
neutral hydrogen. Combined with the measured outflow velocities of
a few hundred $\kms$, this is an indication for energetic winds
that persist for a few hundred million years and drive away nearby 
intergalactic gas.  A strong observed correlation of galaxies with
intergalactic metals supports the idea that the IGM has been enriched
by the outflows from Lyman-break galaxies.
In several cases,
the outflow rate is observed to be proportional to the star formation rate
(Martin 1999a),
consistent with a stellar feedback origin for the outflows.
This body of evidence indicates that supernova-driven winds actually
exist, which helps motivating our theoretical modeling of the relevant
features of galaxy formation.

In this paper we improve the DS scenario for the formation of dwarf galaxies
in view of the developments in galaxy formation theory 
and the refined observed scaling relations across the 
LL/dwarf 
family.
Using a simple energetics criterion and standard assumptions regarding
the origin of galaxy sizes, we now show that the observed scaling relations  
naturally {\it emerge} from the {\it simplest} possible supernova feedback 
scenario, even before one tries to model and simulate in detail the complex 
physics of the feedback mechanism, and before one worries about 
the different types of dwarf and LSB galaxies and the scatter in their 
properties. 
We then address the possible role of radiative feedback in distinguishing 
between dE and dI galaxies, 
and in preventing star formation altogether in some halos. 

\medskip
In \se{general} we address the role of standard assumptions in determining
the scaling relations for galaxies in general.
The assumptions include spherical collapse to
virial equilibrium in CDM haloes and angular-momentum conservation.
In \se{HSB} we apply the analysis to
bright galaxies where we assume that $\Ms/M \simeq \const$. 
In \se{SN} we summarize the DS derivation of the velocity characterizing
  the supernova-feedback scale.
In \se{LSB} we use simple theoretical considerations regarding supernova
feedback to derive the scaling relations of the 
LL/dwarf 
family.
In \se{LSB_LG} we compare the model predictions to the observed relations
shown by the Local Group dwarfs, and comment on the comparison with the SDSS
data.
In \se{rad} we discuss the possible role of radiative feedback in 
distinguishing between dEs and dIs. 
In \se{conc} we discuss our results and related issues.

\section{Scaling Relations: General}
\label{sec:general}

We show that the basic observed scaling relations for galaxies in the two 
regimes can be reproduced to a surprising accuracy based on the 
simplest possible physical assumptions.
These include the {\it virial} theorem for {\it spherical}
{\it cold-dark-matter} haloes, and the notion that a fraction $\eta$ 
of the original gas makes stars in a disc such that the size 
of the stellar system is determined by {\it angular momentum}.
For HH galaxies we recover the scaling relations by taking $\eta$ to
be independent of halo mass, assuming that feedback is not too effective 
there. 
For 
LL 
galaxies, where feedback is a key factor, we use in \se{LSB} below  
a simple energy constraint for the efficiency of supernova feedback 
to predict how $\eta$ should vary with the halo virial velocity. Together
with the constraints from the virial theorem and angular momentum, this
leads to the characteristic scaling relations in the 
LL 
regime.
In this section we derive the scaling relations in general terms
without specifying the behavior of $\eta$.

Assume a dark-matter halo of mass $M$ reaching virial equilibrium
at a time corresponding to cosmological expansion factor $a=(1+z)^{-1}$.
The virial radius $R$ is defined in the spirit of the spherical collapse
model by a given density contrast $\Delta$ relative to
the mean universal density at that time,
namely by $M/R^3 \prop \Delta\, a^{-3}$.
At early times, when $\omm \simeq 1$, the relevant density contrast is
$\Delta \simeq 180$,
while for the standard $\Lambda$CDM cosmology (with $\Omega_\Lambda=0.7$
and $\omm =0.3$ today) it rises to $\Delta \simeq 340$ today.
In the following, we ignore the weak redshift dependence of the $\Delta$ 
factor.\footnote{The maximum change is obtained at low
redshifts. For example, in the range $z=0-2$ the change is
roughly $\Delta \prop a^{1/2}$, which implies that $a$ in the following
expressions should be replaced by $\Delta^{-1/3} a \prop a^{5/6}$.
This is a weak effect, which becomes even weaker at higher redshifts.}
The virial velocity is defined by the virial theorem, $V^2 \prop M/R$, such
that the three virial quantities at $a$ define a one-parameter family:
\be
M \prop a^{3/2} V^3 \prop a^{-3} R^3 .
\label{eq:vir_a}
\ee

In the simplest analysis we ignore the possible systematic increase of $a$
as a function of halo mass (but see below). This dependence is relatively weak
already as predicted by cosmological spherical collapse in the
$\Lambda$CDM cosmology, it gets weaker for smaller haloes as the rms 
density fluctuations approach a constant on small scales,
and it is weaken further by effects like the merging of early-forming small
haloes into bigger ones (see Wechsler \etal 2002).
The virial relations for typical haloes thus take in this approximation
the simple form 
\be
M \propto V^3 \prop R^3.
\label{eq:vir}
\ee
We may keep tracing the $a$ dependence in the general expressions 
below in order to allow small corrections due to its 
possible weak dependence on $M$, when desired.

Considering next the baryonic component,
we assume that the halo is initially filled with gas of mass
$\Mg \sim \fb M$, where $\fb$ ($\simeq 0.13$) 
is the universal baryonic fraction.
For large galaxies,
the gas is assumed to be shock-heated to the halo virial temperature, but
as long as $M < 10^{12}-10^{13}\msun$ the gas in the halo can cool
in a dynamical time (shorter than the Hubble time) and contract to form stars
(Rees \& Ostriker 1977; Silk 1977; White \& Rees 1978; Blumenthal \etal 1984
in the context of dark haloes).
We denote the ratio of luminous stellar mass $\Ms$ to 
initial gas mass $\Mg$ by $\eta$,
\be
\Ms \equiv \eta \Mg \prop \eta M ,
\label{eq:eta1}
\ee
without yet specifying how $\eta$ may depend on $M$.
Substituting \equ{eta1} in the virial relations, \equ{vir_a},
we straightforwardly obtain a general TF relation between $V$ and $\Ms$:
\be
V \prop a^{-1/2} \eta^{-1/3} \Ms^{1/3} .
\label{eq:TF}
\ee
As long as the halo rotation curves are roughly flat at large radii, we ignore
the difference between the virial velocity $V$ and the observed velocity
$\Vmax$.

If the baryons within the halo virial radius $R$ cool and
contract to a centrifugally supported disc of radius $\Rs$ while
preserving their specific angular momentum $j$, then, following
Fall \& Efstathiou (1980) and Mo, Mao \& White (1998), we write
$\Rs \simeq \lambda R$, where $\lambda = j / (RV)$
is the initial baryonic spin parameter (according to the revised, 
practical definition of Bullock \etal 2001b).
Then, from the virial relations above,
\be
\Rs \prop \lambda a M^{1/3} .
\label{eq:Rs}
\ee
With $\Ms \prop \eta M$, this implies for the surface brightness
\be
\mus \prop \Ms \Rs^{-2}
\prop \lambda^{-2} a^{-2} \eta^{2/3} \Ms^{1/3} . 
\label{eq:mus}
\ee

The characteristic radii and surface brightnesses
derived for discs can be argued to be roughly valid also for the 
spheroidal stellar components, elliptical galaxies or bulges of spirals.
This is based on energy conservation and the virial theorem
under the assumption that the spheroids were formed by mergers of discs.

The distribution of halo spin parameter is known from cosmological
simulations to be insensitive to halo mass (Barnes \& Efstathiou 1987; see
Bullock \etal 2001b for the case of $\Lambda$CDM cosmology), 
so for a crude approximation it could be dropped from the
mean scaling relations for the radius and surface brightness (but see a comment
in \se{LSB}).
However, we note in passing that the $\lambda$ dependence in the expressions
for the radius and surface brightness, as opposed to 
its absence from the TF relation, may have an important implication
on the scatter about these relations.
Cosmological simulations of $\Lambda$CDM reveal that the distribution
of halo spin parameter is log-normal with a mean of 
$\lambda = 0.035 \pm 0.005$ and a standard deviation in the (decimal) log of
$\sigma_{\lambda} = 0.184 \pm 0.011$ (Bullock \etal 2001b).
This is already comparable to the standard deviation of the distribution
of stellar radii for the 
LL 
galaxies in SDSS (K03, Fig.~8),
even before considering the additional scatter in $a$ and in $\eta$.
The absence of $\lambda$ dependence in the TF relation, and its weaker
dependence on $a$ and $\eta$, may explain why the TF relation is much 
tighter (see Courteau \etal 2003).

The amount of metals produced in a galaxy is assumed to be proportional to 
$\Ms$ with a constant yield $y$. When $\eta \simeq 1$, the metallicity is simply
$Z \simeq y$. When $\eta \ll 1$, in the instantaneous-recycling approximation
(e.g., Searle \& Sargent 1972; Audouze \& Tinsley 1976), one expects 
$Z \sim y \eta$. So we approximate in general 
\be
Z \prop \Ms/\Mg = \eta .
\label{eq:Z}
\ee

As mentioned above,
small corrections to the above relations may result from a
correlation between the time of formation of a halo and its mass.
An upper bound to this effect may be obtained from the straightforward
prediction based on spherical collapse, ignoring the fact that many
early-forming small haloes eventually merge to bigger ones and thus  
weaken the $a(M)$ relation.
For a power spectrum of linear density fluctuations that resembles
the power law $P_k \prop k^n$ at the vicinity of the scales relevant
for galactic haloes, the typical mean density fluctuation within a proto-halo
is $\delta \prop M^{-(n+3)/6} D(t)$,
where $D(t)$ is the linear growth rate, $D(t) \prop a$ for the
Einstein-deSitter cosmology relevant at high redshifts.
The formation time in the spherical collapse model can be approximated by
$\delta \simeq 1.7$ for the linearly extrapolated mean density fluctuation, 
so one obtains
\be
a \prop M^{(n+3)/6}.
\label{eq:a_n}
\ee
The virial relations for typical haloes, \equ{vir_a}, thus become
\be
M \propto V^{12/(1-n)} \prop R^{6/(5+n)} .
\label{eq:vir_n}
\ee
The TF relation, \equ{TF}, is now  
\be
V \prop (\eta^{-1}\Ms)^{(1-n)/12} .
\label{eq:TF_n}
\ee
The stellar radius is now given by $\Rs \prop \lambda M^{(5+n)/6}$ such that
the surface brightness, \equ{mus}, is replaced by
\be
\mus \prop \lambda^{-2} \eta^{(5+n)/3} \Ms^{-(2+n)/3} .
\label{eq:mus_n}
\ee

The expressions so far should be valid in general,
both for 
HH and LL 
galaxies. 
The differences between the two classes enter mainly via the behavior of
$\eta$, with an additional weak effect due to the difference in the effective
$n$ in the maximum $M$ dependence of $a$.

\section{HH Galaxies}
\label{sec:HSB}

For HH galaxies we take $\eta$ to be roughly independent of halo mass.
This is based on the assumption that feedback effects do not significantly
heat or remove most of the gas from these galaxies such that most of the gas,
or a constant fraction of it, eventually forms stars (see \se{SN}).
With $\eta$ and $a$ independent of mass in \equ{TF}, \equ{mus} and \equ{Z},
the scaling relations for HH galaxies become 
\be 
V \prop \Ms^{1/3}, \quad
\mus \prop \Ms^{1/3}, \quad
Z\simeq {\rm const.}
\label{eq:scaling_hsb}
\ee
These are already in qualitative
agreement with the observed relations for HH galaxies.

When considering the limit of maximum $M$ dependence of $a$,
we recall that
big galactic haloes in a $\Lambda$CDM cosmology correspond to the part of
the power spectrum where $n \lsim -2$. For example, with $n=-2$ at the 
bright end, one has $a \prop M^{1/6}$.  
Then the virial relations become $M \propto V^4 \prop R^2$.
With $\eta$ assumed independent of $V$ for HH galaxies,
and with the maximum $M$ dependence of $a$ computed above,
the predicted scaling relations become
\be
\Vmax \prop \Ms^{1/4}, 
\quad \mus \simeq \const,
\quad Z \simeq \const 
\label{eq:scaling_n_hsb}
\ee 
In order to compare with observations in terms of luminosity rather 
than stellar mass, 
one can assume 
that for HH galaxies the stellar mass-to-light ratio varies like 
$\Ms/L \prop L^{0.3}$
(e.g., Courteau \etal 2003). 
[This is based, for example, on the reading of Fig.7 of Courteau \etal 
that $(V-I) \simeq -0.09 M_I$, combined with the result from Table 1 of 
Bell \& de Jong (2001) that $\log(\Ms/L_I) \simeq 1.35 (V-I)$. 
A similar result is obtained 
for elliptical galaxies (Bender, Burstein \& Faber 1992).] 
With $\Vmax \simeq V$
we thus roughly recover the observed TF relation: $L \prop \Vmax^3$. 
The surface brightness as measured in terms of luminosity 
is predicted to be slowly decreasing with luminosity: 
$I \prop L/\Rs^2 \prop L^{-0.3}$, in qualitative agreement with observations.
We note that in the TF relation
the correction due to the correlation of $a$ and $M$
roughly balances the correction due to the correlation of $\Ms/L$ and $M$.
However, the corresponding corrections to the relation
of surface brightness and luminosity add up.
In any case, the corrections to the simple predictions of
\equ{scaling_hsb} are small, and we expect the 
predicted scaling relations for HH galaxies to lie somewhere between
the relations in \equ{scaling_hsb} and \equ{scaling_n_hsb}. 
This range is in general
agreement with the observed scaling relations for HH galaxies. 

Recall that, beyond the standard assumptions of virial equilibrium
and spherical collapse, the key assumption for HHs was that most of 
the original gas, or a constant fraction of it, turns into stars, 
namely, $\eta \simeq {\rm const.}$

\section{Supernova Feedback: the Critical Scale}
\label{sec:SN}

Dekel \& Silk (1986) evaluated the maximum total energy fed into the
interstellar gas by a
collection of supernova explosions due to a period of star formation at
a constant rate $\sfr$, taking into account the radiative loses
(based on the standard evolution of a supernova remnant in a uniform
interstellar medium, e.g. Spitzer 1978; Ostriker \& McKee 1988).
They found that at time $t$ this energy can be approximated by
\be
\Esn(t) \simeq \epsilon \nu \sfr \trad f(t) ,
\label{eq:ds_44}
\ee
where $\epsilon$ is the initial energy released by a typical supernova
($\epsilon \sim 10^{51} {\rm erg}$)
and $\nu$ is the number of supernovae per unit mass of forming stars
(which for a typical IMF is $\nu \sim 1$ per $50 \msun$ of stars).
The characteristic time $\trad$ marks the end of the ``adiabatic" phase
and the onset of the ``radiative" phase of a typical supernova remnant,
by which it has radiated away a significant fraction of its energy.
The dimensionless factor $f(t)$ turns out to be of order unity when
$t \sim \trad$;
it grows roughly $\prop t$ for $t<\trad$ and $\prop t^{0.4}$ for $t>\trad$.

We assume here that the stellar population of mass $\Ms$ has formed
over some constant multiple $\tau$ of the free-fall time $\tff$, namely
\be
\sfr = {\Ms \over \tau \tff}.
\label{eq:sfr} 
\ee 
Substituting in \equ{ds_44} we obtain that the total energy fed into the gas is
\be
\Esn \prop \Ms {\trad \over \tff} .
\label{eq:Esn_t}
\ee 
DS noticed that in the temperature range $6\times 10^4<T<6\times 10^5$K
the cooling rate scales approximately like $\Lambda \prop T^{-1}$,
which implies that the ratio $\trad/\tff$ is roughly a constant,
of order $10^{-2}$, independent of the gas density or the halo parameters.
This leads to
$\Esn \propto \Ms$,
which we show below is a key for deriving the scaling relations
of 
LL 
galaxies (\se{LSB}).
Note that DS originally assumed $\sfr \prop \Mg/\tff$ rather than
the $\prop \Ms/\tff$ of \equ{sfr}. The two assumptions are roughly
equivalent in the case of bright
galaxies and when trying to estimate the transition scale between 
HH and LL 
galaxies where $\Ms \sim \Mg$.

DS also showed that if star formation is rapid, $\tau \sim 1$,
then the filling factor of the expanding supernova shells within the halo
is of order unity when the typical shell is at
the end of its adiabatic phase, at $\trad$.
This coincidence indicates that the supernova energy (minus the radiative
losses) can be fed quite evenly and efficiently into most of the gas
via the expanding shells that reach a significant mutual overlap roughly at
the time after which they become ineffective.
It also justifies the adoption of $f\sim 1$ in \equ{ds_44}.

A necessary condition for heating or unbinding most of the initial gas of
mass $\Mg$
is obtained by requiring that the energy fed by supernovae is comparable to the
binding energy of the gas in the halo potential well,
\be
\Esn = (1/2) \Mg V^2 .
\label{eq:MgV2}
\ee
Here $V$ is the virial velocity of the halo, which we assume for simplicity
to be isothermal and to dominate the potential.
DS then pushed this approximate relation to the limit where a large fraction
of the gas turns into stars and obtained the critical velocity
\be
\Vsn =  \left( 2 f \epsilon {\nu \over \tau} {\trad \over \tff} \right)^{1/2}
\simeq 100\kms .
\label{eq:Vsn}
\ee
This critical velocity is evaluated using the typical values of $\epsilon$ and
$\nu$ with $f \simeq \tau \simeq 1$ and is independent of the gas density
because of the robustness of $\trad/\tff$.
The interpretation of this critical velocity is that gas removal becomes
possible in haloes with virial velocities smaller than $\Vsn$.
We note that the corresponding virial mass is
\be
\Msn \simeq 2.2 \times 10^{11} \msun \left( {\Vsn \over 100 \kms} \right) ^3
a^{2/3} .
\label{eq:Msn}
\ee
With $\Ms \simeq \Mg \simeq \fb M$ and the universal baryonic fraction
$\fb \simeq 0.13$ the corresponding characteristic stellar mass today is
\be
\Mssn \simeq 3 \times 10^{10} \msun ,
\ee
in excellent agreement with the transition scale seen in the SDSS data.

The haloes of bright galaxies, which have retained most of their gas,
are thus limited to the regime of deep potential wells, $V>\Vsn$.
We associate the galaxies that form in haloes below the critical supernova 
scale with 
LL 
or dwarf galaxies. The importance of feedback effects in the
history of these galaxies implies that their scaling relations can be
very different from those shown by galaxies that live in haloes
of virial velocities larger than $\Vsn$.

\section{LL Galaxies}
\label{sec:LSB}

We now use the feedback energetics constraints, \equ{Esn_t} and \equ{MgV2},
to determine the behavior of $\eta$ in the 
LL 
regime,
where we expect feedback effects to allow only a fraction
of the gas to turn into stars, 
\be
\eta \equiv {\Ms \over \Mg} < 1 .
\label{eq:eta}
\ee
The supernovae resulting
from the first burst of stars either blow out the rest of the gas, or at least
provide enough feedback energy to regulate the subsequent star formation
rate and keep it low.
We assume that $\eta \simeq 1$ at $V=\Vsn\sim 100\kms$ 
and that $\eta$ becomes gradually
smaller for haloes of smaller velocities.
The following simple analysis is actually carried out 
in the limit of strong feedback, $\eta \ll 1$.

Our key starting point is \equ{Esn_t} with $\trad/\tff={\rm const.}$, namely
the energy fed into the interstellar gas
by supernovae is proportional to the final stellar mass,
\be
\Esn \propto \Ms .
\label{eq:Esn=Ms} 
\ee 
Without the radiative losses of the supernova energy, this would have been
anybody's first intuitive guess for a relation between these quantities.
We argue here, in the spirit of the DS analysis, that the actual energy fed 
into the gas after significant radiative losses is still a
constant fraction of the original supernova energy. 
This makes \equ{Esn=Ms} a valid approximation in the realistic case.

In order to allow significant heating or total blowout of the initial gas, 
the total input by supernovae should be at least comparable 
to the binding energy of the gas,  \equ{MgV2}.
With \equ{Esn=Ms}, the energy condition becomes $\Ms \prop \Mg V^2$, namely
\be
\eta \prop V^2 .
\label{eq:energy}
\ee
The scaling relations for 
LLs 
all follow from this basic relation,
which measures the strength of the feedback effects 
along the halo sequence characterized by the parameter $V$
in the range $V<\Vsn$.

\Equ{energy}, combined with the virial relations for the halo, 
\equ{vir_a}, and then $\Ms = \eta M$, yield 
\be
\eta \prop a^{-1} M^{2/3}
\prop a^{-3/5} \Ms^{2/5} .
\label{eq:eta_Ms}
\ee
Recall that in the instantaneous-recycling approximation, for $\eta \ll 1$,
the metallicity is simply 
\be
Z \prop \eta ,
\label{eq:Z_LSB}
\ee
so the mean scaling relation involving metallicity is given by \equ{eta_Ms}.

Substituting $\eta$ in \equ{TF} we obtain for the TF relation in the 
LL 
regime 
\be
V \prop a^{-3/10} \Ms^{1/5} .
\label{eq:TF_LSB}
\ee

Then substituting $\eta$ in \equ{Rs} and \equ{mus} we obtain for the radius
$\Rs \prop \lambda a^{6/5} \Ms^{1/5}$
and for the surface brightness
\be
\mus \prop \lambda^{-2} a^{-12/5} \Ms^{3/5} .  
\label{eq:mus_LSB}
\ee

In summary, 
when ignoring possible weak systematic dependences of $a$ and $\lambda$ 
on $M$, the scaling relations for 
LL/dwarf 
galaxies are predicted to be 
\be
V \prop \Ms^{1/5}, \quad Z \prop \Ms^{2/5}, 
  \quad \mus \prop \Ms^{3/5} .
\label{eq:scaling_LSB}
\ee

In order to evaluate the maximum correction due to the possible
dependence of $a$ on $M$, we use \equ{a_n} in \equ{eta_Ms} and obtain
\be
\eta \prop M^{(1-n)/6} \prop \Ms^{(1-n)/(7-n)} .
\label{eq:eta_n}
\ee
Then the TF relation, \equ{TF_LSB}, becomes
\be 
V \prop a^{-3/10} \Ms^{1/5} \prop \Ms^{(1-n)/(14-2n)} .
\label{eq:TF_LSB_n}
\ee
and the surface brightness, \equ{mus_LSB}, becomes
\be
\mus \prop \lambda^{-2} \Ms^{-3(1+n)/(7-n)} .
\label{eq:mus_LSB_n}
\ee
Very small galaxies in the $\Lambda$CDM cosmology correspond to the part of 
the power spectrum where $n$ is not much larger than the lower limit of
$n=-3$, implying a similar formation time for dwarf galaxies of all masses
and therefore a constant $a$ in the above relations, thus leading to 
\equ{scaling_LSB}. 
For 
LL 
galaxies not much below $\Vsn$ we may
try for example a typical $n=-2.5$, for which 
$a \prop \Ms^{(n+3)/(7-n)} \prop \Ms^{1/19}$.
This implies negligible effects on the TF relation and the metallicity 
relation, but the weak correction to the surface brightness relation
may be marginally detectable, $\mus \prop \lambda^{-2} \Ms^{9/19}$
compared to $\mus \prop \lambda^{-2} \Ms^{3/5}$. 

In this case, however, we may also wish to 
incorporate the possible mass dependence of $\lambda$.
To a first approximation, as said above based on cosmological simulations,
the distribution of halo spin parameter is independent of the halo virial 
properties and its formation time. As long as the baryons initially
trace the spatial distribution and kinematics of the halo, their
$\lambda$ distribution can be assumed independent of $M$ and $a$.
However, while the baryons in bright disc galaxies seem to have spin
parameters similar to those of their host haloes, 
LSB disc galaxies may tend to be associated with a higher spin
parameter.  For example, van den Bosch, Burkert \& Swaters (2001,
hereafter BBS) studied the spin in
a sample of 14 LSB discs with an estimated average
of $V \simeq 60\kms$. They found an average spin parameter about 50\% 
larger than that of the dark haloes (see Maller \& Dekel 2002, Fig.~8). 
At the same time, BBS estimated in these galaxies an average baryonic fraction 
of only $\fd \simeq 0.035$, which translates in our terminology to
$\eta(V=60) = \fd/\fb \simeq 0.27$. 
Following Maller \& Dekel (2002), we model 
these systematic trends based on preferential blowout of low-spin material
in dwarf galaxies. 
In order to obtain the 50\% change in spin parameter over the
same $\eta$ range (between 1 and 0.27) 
the effect of blowout should roughly scale like 
$\lambda \prop \eta^{-0.3} \prop a^{0.18} \Ms^{-0.12}$.  
Plugged into \equ{mus_LSB}, using $n=-2.5$, we now obtain
$\mus \prop \Ms^{0.7}$.
This kind of correction to the surface-brightness relation should be valid for 
relatively large 
LL 
galaxies, where $n$ is not too close to $-3$ and where 
the BBS analysis indicates a systematic spin dependence. 
For smaller dwarfs the actual 
relation may be better approximated by \equ{mus_LSB} with constant $a$ and
$\lambda$, namely $\mus \prop \Ms^{0.6}$.

The scatter about the mean scaling relations is expected to partly reflect the
random scatter about the mean $a$ and $\lambda$.
Based on \equ{mus_LSB}, the scatter about the mean relation $\mus(\Ms)$ is
expected to be significant, dominated by the scatter in $\lambda$, 
while the scatter in $V$ and $Z$ is expected to be smaller.
The residuals in these different relations are expected to be correlated.
For a given $\Ms$, galaxies that lie at the bottom of the
$\mus$ distribution are expected to be of relatively high $a$ 
(late formation time) 
and high $\lambda$. In turn, based on \equ{TF_LSB} and \equ{Z_LSB}, 
these galaxies compared to the average for that $\Ms$ are expected 
to be of low $V$ (though high $M$, given the high $a$) and low $Z$.

\section{Model vs. Local-Group Dwarfs}
\label{sec:LSB_LG}

The success of the simple feedback model for 
LL 
galaxies as described in \se{LSB} can 
be evaluated by comparing the predicted scaling relations, \equ{scaling_LSB},
to the observed scaling relations for 
LLs 
in SDSS and  
in the Local Group. Given the idealized nature of the straightforward model,
one might only hope for a crude qualitative fit.

The match of the predicted characteristic
scale for supernova feedback, $V \simeq 100\kms$,
with the observed transition at $\Ms \simeq 3\times 10^{10}\msun$ is
already remarkable; it indicates that this transition may indeed be associated
with the onset of supernova feedback effects. 

\begin{figure}
\vskip 8.0cm
{\includegraphics{f1.eps}}
\caption{
Central surface brightness versus stellar mass for the Local Group dwarfs
(from WD).
Shown are the regression line $\mus \prop \Ms^{0.55}$ (solid), 
the correlation coefficient $r$, and the toy-model theoretical prediction
$\mus \prop \Ms^{0.6}$ normalized for best fit (dashed).
}
\label{fig:mu}
\end{figure}

\Fig{mu} shows the central surface brightness $\mus$ versus stellar mass $\Ms$
for the Local Group dwarfs (from WD).
The galaxies are either of the two major types, dI and dE, 
or transition cases marked Tr.
The data are fit very well by the predicted scaling relation 
$\mus \prop \Ms^{0.6}$ throughout the whole 
LL 
range,
spanning 5 decades in $\Ms$. We do not attempt to normalize the predicted
relation, and therefore the model line in the figure is normalized artificially
to provide the best fit for the predicted slope of 0.6.
The correlation is relatively tight, with a Pearson's
correlation coefficient for the logs of $r=0.88$.
The model slope is also a good fit to the SDSS data in the 
LL 
range (Kauffmann \etal 2003b); even the predicted slight steepening 
to $\mus \prop \Ms^{0.7}$ or so can be seen at the bright end of the 
LL 
range.
The SDSS data refers to the surface brightness within the half-light radius,
which, for an exponential profile, is a factor of $\sim 3$ 
smaller than the central value. With this relative normalization, the
bright end of the Local-Group dwarfs lies along the upper 68\% 
contour of the SDSS distribution (Fig.~7a of Kauffmann \etal 2003b).

\begin{figure}
\vskip 8.0cm
{\includegraphics{f2.eps}}
\caption{
Metallicity versus stellar mass for the Local Group dwarfs (WD).
Shown are the regression line $Z \prop \Ms^{0.40}$ and the toy-model
theoretical prediction $Z \prop \Ms^{0.4}$ (normalized for best fit).
}
\label{fig:Z}
\end{figure}

\Fig{Z} shows the metallicity $Z$ versus stellar mass $\Ms$
for the Local Group dwarfs (WD).
The predicted metallicity relation, $Z \prop \Ms^{0.4}$, is a very good
fit to the Local Group dwarfs. The correlation is tight, with
a correlation coefficient for the logs of $r = 0.92$.
The preliminary SDSS data indicate a similar
and perhaps slightly steeper relation at the bright end, $Z \prop \Ms^{0.5}$.

\begin{figure}
\vskip 8.0cm
{\includegraphics{f3.eps}}
\caption{
Velocity versus stellar mass for the Local Group dwarfs (WD).
Shown are the regression line $V \prop \Ms^{0.24}$ over the whole range
and the toy-model theoretical prediction $V \prop \Ms^{0.2}$.
Note the lower bound at $V \simeq 10\kms$ for the dEs below $3\times10^7\msun$.
}
\label{fig:V}
\end{figure}

\Fig{V} shows the velocity $V$ versus stellar mass $\Ms$
for the Local Group dwarfs (WD).
The predicted relation, $V\prop \Ms^{0.2}$, is an acceptable eye-ball fit
to the data, despite the fact that the formal regression slope is 
somewhat steeper, $0.24 \pm 0.01$.
The correlation is tight, with $r=0.89$.

We see that the idealized theory for supernova feedback
provides a surprisingly good fit to the characteristic scale and to the
three independent scaling relations valid across the whole 
LL 
range.
This indicates that the supernova feedback effects, via the parameter
$\eta$, indeed have a primary role 
in determining the gross features of the galaxy properties in the 
LL 
regime.

The data from SDSS also allow a quantitative evaluation of the distribution
of galaxies about the mean relation in the $\mus$-$\Ms$ plane 
(Kauffmann \etal 2003b). 
In the 
LL 
regime, the spread in $\mus$ at a given $\Ms$
is roughly consistent with the spread in spin parameter $\lambda$ for haloes
of a given mass as measured in N-body simulations of $\Lambda$CDM.
This indicates that $\lambda$ can indeed serve as the main secondary parameter 
for the 
LL 
family.

The idealized theory predicts a dependence of $\mus \prop \lambda^{-2}$
for a given $\Ms$ [\equ{mus}]. In \Fig{mu_lambda} we test the self-consistency 
of this prediction for the Local Group dwarfs. 
We display $\mus$ versus an estimated
``spin parameter" given by $\lambda \prop \Rs/R$, the ratio
of stellar radius to halo radius.
The stellar radius is determined from $\Ms$ and the
central surface brightness $\mus$, $\Rs \prop (\Ms/\mus)^{1/2}$. 
The halo radius is the virial radius
corresponding to virial velocity $V$, where $V$ 
is $\max\{V_{\rm circ},\sqrt{3}\sigma_{\rm p}\}$ as described in WD.
When dividing the galaxies into three relatively narrow bins of $\Ms$ values,
we see that there is indeed a systematic trend within each bin, 
though slightly flatter than the expected $\mus \prop \lambda^{-2}$.

\begin{figure}
\vskip 8.0cm
{\includegraphics{f4.eps}}
\caption{
Surface brightness versus ``spin parameter" in three bins of constant
$\Ms$ values.
The ``spin parameter" is actually the ratio of stellar radius to halo radius,
$\lambda \prop \Rs/R$. The decreasing trend is to be compared to the predicted
$\mus \prop \lambda^{-2}$, \equ{mus}.
}
\label{fig:mu_lambda}
\end{figure}

Although the surface brightnesses of dEs and dIs follow in general a 
similar scaling relation,
the dIs do tend to lie somewhat below the best fit line.
This is consistent with the finding in SDSS that at a fixed $\Ms$
the galaxies with lower bulge-to-disc ratio and younger stellar populations
tend to have a lower surface brightness.
These trends are qualitatively consistent with the $a$ and $\lambda$ 
dependences predicted in \equ{mus_LSB}. 

While the predicted $V \prop \Ms^{0.2}$ is a good fit across the whole dwarf
range, a more detailed investigation of \Fig{V} reveals very interesting
secondary features.
First, there is an apparent lower bound for galaxies at $V \simeq 10\kms$.
Second, there is an apparent transition at $\Ms \simeq 3\times 10^7\msun$. 
The fainter galaxies can actually be well fit by $V \simeq {\rm const.}$.
The dwarfs brighter than $3\times 10^7\msun$ are then fit by 
a line which could be as steep as $V\prop \Ms^{0.4}$. 
Since these velocities are measured in the inner regions of the haloes,
they can be regarded as lower bounds to the actual dispersion velocities
of the haloes. If the velocities of the dE haloes are actually larger 
than these lower estimates (as argued by Stoehr \etal 2002), then the 
difference between the TF relation in the two regimes, below and above
$3\times 10^7\msun$, could become even more significant.
We note that the faint part is dominated by dEs while the brighter part
is mostly dIs. These are clues for the origin of the distinction between
these two types of dwarf galaxies, which we address in the following section.

\section{Radiative Feedback: {$\pmb\d$}E vs. {$\pmb\d$}I }
\label{sec:rad}

After demonstrating the encouraging success of supernova feedback 
in explaining the
basic systematic trends in the 
LL 
family as a whole, we now attempt to
consider the possible role of another feedback mechanism,
and in particular how it may differentiate between dEs and dIs 
within the 
LL 
family.
 
\subsection{Radiative feedback}

Cosmological reionization of Hydrogen is complete by $\zion \sim 6-7$ 
(see a review by Barkana \& Loeb 2001, hereafter BL; 
also Loeb \& Barkana 2001).
The flux of UV radiation that is generated by the first stars or AGNs
heats and photoionizes the gas in the IGM and in virialized haloes
(except perhaps for the inner regions which can become shielded).
As long as the ionizing flux persists, the gas is kept 
at a fixed temperature of $\Tion \simeq (1-2)\times 10^4$K. 
This can be regarded as another feedback mechanism; it
can suppress star formation and clean haloes from gas in two ways.  
First, by photo-evaporation of gas already in haloes.
Barkana \& Loeb (1999) estimated that haloes of $V < 10\kms$ would
lose most of their gas. Their analysis provides an estimate of
the gas loss during the first dynamical time after $\zion$.
However, if the gas is kept ionized until $z\sim 1-2$, 
a dynamical calculation of evaporation by a continuous wind reveals
that photo-evaporation would remove the gas from 
somewhat larger haloes, up to $\Vev \simeq 20\kms$ (Shaviv \& Dekel 2003).
Second, based on simulations and computations of Jeans mass, 
the pressure of the hot IGM shuts off gas infall into even more massive haloes,
those with velocities up to $\Vjeans \simeq 30\kms$ 
(see BL \S 6.5 and references therein).
Gas could resume falling into small haloes after $z \sim 1-2$ when the UV
background flux declined sufficiently (Babul \& Rees 1992), but only
haloes of $V>20-25\kms$ can form molecular hydrogen by $z\sim 1$ and then
cool further to make stars (Kepner \etal 1997).

In the presence of a halo potential well characterized by a velocity $V$,
the fraction of gas of temperature $\Tion$
that is bound to the halo can be estimated by the Boltzmann distribution,
\be
\fbound \propto 1 - e^{(-V^2/k\Tion)}.
\label{eq:Boltz}
\ee
The velocity corresponding to $\Tion$ is on the order of $\Vev \sim 20\kms$ 
mentioned above.
Note that in the limit $V<\Vev$, \equ{Boltz} predicts $\fbound \prop V^2$
(as pointed out by J.P. Ostriker in a private communication).
This reminds us of the energy relation for supernovae, $\eta \prop V^2$,
which led to the global scaling relations of 
LLs 
in \se{LSB}.
Could radiative feedback (rather than supernova feedback)
be the actual mechanism responsible for the global scaling relations of 
LLs? 
First, it is unlikely that a mechanism whose characteristic scale is 
$\sim 20-30\kms$ can be dominant in determining the observed critical 
scale of $\sim 100 \kms$ and the properties of bright 
LLs 
not much below this scale.
Second, $\fbound$ in the radiative case refers to the sum of bound mass 
in stars and in gas while $\eta$ in the supernova case refers to the 
stellar mass only.
While $\Ms /M \prop V^2$ is consistent with the observed scaling
relations for 
LLs, 
a similar relation for the gas-to-mass ratio 
does not seem to be in agreement with the observed trend, especially not 
in the large, gas-rich 
LLs. 
There are indications that as one moves from bright to fainter 
galaxies the ratio of gas to
stellar mass increases until it reaches a maximum at some intermediate
scale typical to dIs before it starts decreasing towards the dE regime
(e.g., McGaugh \& de Blok 1997).
We interpret this as another evidence 
against radiative feedback being the dominant
mechanism in determining the global properties of galaxies in the 
upper 
LL 
regime.  
The radiative feedback should have an important effect though
in the small, gas-poor
dEs, and possibly a complementary effect to the supernova feedback
in the larger, gas-rich dIs and 
LLs. 

In haloes of $V<\Vev$, stars can form only before the reionization epoch
(and possibly much later, at $z<1-2$).
If $\Vev<V<\Vjeans$, gas that cooled and collapsed before the reionization
epoch can turn into stars in a slow rate also at later times, 
but new gas cannot be accreted. These effects could lead to the gas-poor dEs.
On the other hand, in haloes of $V > \Vjeans$ there is no much radiative gas 
loss.  Galaxies that form in such haloes can retain some gas that has not 
been blown away by supernova winds, or has come back after such blowout, 
and thus give rise to gas-rich dIs.
We thus propose that the main role of radiative feedback is to clean up the 
dEs from their gas and to help regulating star formation in dIs.

\subsection{Dwarf elliptical galaxies}
\label{sec:dE}

\Fig{Vsplit} shows the same data as in \Fig{V}, but with separate fits
below and above $\Ms=3\times 10^{7}\msun$, in the ranges dominated by
the dwarf spheroidals of the Local Group and by dIs respectively. 
In the low-$\Ms$ range we fit a horizontal line, $V = 15\kms$, and
then determine the best-fit slope in the high-$\Ms$ range, 
$V \prop \Ms^{0.37}$. 
This fit with a broken line naturally provides a better fit than 
with a single line across the whole 
LL 
range.
It does provide a crude hint for different TF relations in the two regimes.

\begin{figure}
\vskip 8.0cm
{\includegraphics{f5.eps}}
\caption{
Velocity versus stellar mass for the Local Group dwarfs (WD), same data
as in \Fig{V}.
Shown here is the best fit horizontal segment below $\Ms=3\times 10^7\msun$
and the best fit line at larger stellar masses: $V \prop \Ms^{0.37}$. 
}
\label{fig:Vsplit}
\end{figure}

We thus propose that haloes in the range $10 < V < 30\kms$ tend to
form gas-poor dEs.
Efficient cooling by Hydrogen recombination at $T\gsim 10^4$K
leads to an early burst of stars.
The associated supernovae blow out much of the gas. The rest of the gas
photo-evaporates (if $V<\Vev$) and no new gas can fall in (if $V<\Vjeans$),
leaving behind a gas-poor system with no significant recent star formation.
The gas in haloes of $V < 10 \kms$ cannot cool to form stars at any early
epoch.
This confines all the dEs to a narrow range of halo velocities, $V \simeq
20\pm10\kms$. The spread of $\Ms$ within the dE family is thus predicted
not to correlate with $V$ (as it does for the dIs) 
but rater to represent variations in other quantities such as
the time available for star formation between the halo collapse and $\zion$ 
(as suggested by Miralda-Escude \& Rees 1998).
The spread in $\Ms$ for a given halo $V$ is large for the dEs and
much smaller for the dIs and the rest of the 
LL 
family (\Fig{V}).
This is because the gas that remains in dI haloes (or falls back in later on)
allows star formation to continue after reionization,
bringing $\Ms$ close to its value predicted by the energy requirement
$\eta \prop V^2$.

If radiative feedback is indeed important in the formation of dEs, we should
try to understand why the scaling relations shown by the dEs for the 
metallicity and surface brightness versus stellar mass seem to be 
natural extrapolations of those shown by the dIs, which we associated
with supernova feedback.
For example, 
if all dEs have similar haloes and therefore similar initial gas masses,
and if the metals were assumed to be uniformly distributed throughout the gas
as assumed for larger 
LLs, 
then we might have expected a steeper dependence of
$Z\prop \Ms/M \prop \Ms$ within the dE range (rather than the global
$Z\prop \Ms^{0.4}$ which applies throughout the 
LL 
range).
However, since some of the gas is expected to photo-evaporate or be kept
away from the halo even before it cools and falls into the halo center,
we expect the metals to enrich a smaller fraction of the initial gas
in fainter dEs, which should lead to a weaker dependence
of $Z$ on $\Ms$, perhaps as flat as $Z\prop \Ms^{0.4}$.
Given the limited width of the dE range,
we can probably tolerate some deviation from the global
$Z\prop \Ms^{0.4}$ there.

\begin{figure}
\vskip 8.0cm
{\includegraphics{f6.eps}}
\caption{
Stellar radius versus stellar mass for the
dEs in the range $\Ms \leq 3\times 10^7\msun$.
Best fit is $\Rs \prop \Ms^{0.3}$.
Shown for comparison is $\Rs \prop \Ms^{0.2}$ (see the text).
}
\label{fig:r_dE}
\end{figure}

As for the surface brightness in dEs,
the similarity in $V$ between all the dE haloes would lead to the
predictions $\Rs \simeq \lambda R \prop \lambda a^{1/2}$
and therefore $\mus \prop \Ms/(\lambda R)^2 \prop \lambda^{-2} a^{-1} \Ms$
(compared to the global $\mus \prop \Ms^{0.6}$).
If the sequence of $\Ms$ in dEs indeed represents variations in formation time,
then the combination of the $a$ and $\lambda$ factors should be 
responsible for the flattening of the $\Ms$ dependence of $\mus$ to the
observed $\mus \prop \Ms^{0.6}$. 
While $a$ is expected to be smaller for larger $\Ms$,
the baryonic spin parameter
is expected to be smaller for dEs of smaller $\Ms$, those
that formed later and closer to $\zion$. This is because in those
only the gas from the inner halo managed to form stars before the
reionization time, and this inner gas is naturally expected to be of lower
than average spin (see Bullock \etal 2001b).
The observed relation of roughly $\mus \prop \Ms^{0.6}$ tells us that
the required trend should roughly be 
$\Rs \prop \lambda a^{1/2} \prop \Ms^{0.2}$.
If $a$ is indeed anticorrelated with $\Ms$ for dEs, we expect for $\lambda$
a stronger dependence than $\lambda \prop \Ms^{0.2}$.
Such a spin gradient could also explain why the dwarfs at the faint end are
low-spin spheroidals while the brighter dwarfs tend to be centrifugally
supported discs.
As a consistency check, 
\Fig{r_dE} shows the stellar radius $\Rs$ versus stellar mass $\Ms$ for the
dEs in the range $\Ms \leq 3\times 10^7\msun$.
There is indeed an apparent trend, best fit by $\Rs \prop \Ms^{0.3}$
and reasonably consistent with the required 
$\Rs \prop \lambda a^{1/2} \prop \Ms^{0.2}$.
Again, given the limited width of the dE range, we can probably tolerate there
a certain deviation from the global relation of $\mus \prop \Ms^{0.6}$.

\section{Discussion}
\label{sec:conc}

We identify four basic characteristic scales in the theory of
galaxy formation, each originating from a different physical process, and each 
having a different imprint on the galaxy population, as follows:
\begin{enumerate}
\item
The upper limit for bright galaxies separating them from clusters of galaxies, 
at $\Ms \sim 10^{12}\msun$,
is where radiative cooling occurs on a dynamical 
time scale (Rees \& Ostriker 1977; Silk 1977; White \& Rees 1978 in 
the context of dark haloes).
The cooling curve in the relevant temperature 
and density range makes this bound roughly coincide with an 
upper bound to the total mass.
\item  
Supernova feedback becomes effective in heating and removing gas 
in haloes of $V<100\kms$, as predicted by Dekel \& Silk (1986). 
We argue that this scale marks the transition between 
HH and LL/dwarf 
galaxies, as seen in the SDSS data near $\Ms \simeq 3\times
10^{10}\msun$. 
\item 
Radiative feedback after $\zion$ heats the gas to $T \sim 10^4$K,
which causes efficient evaporation from haloes of $V<20\kms$ 
(Shaviv \& Dekel 2003) and prevents
further infall into haloes of $V<30\kms$ 
(see a review by Barkana \& Loeb 2001).
We propose the possibility that this scale, corresponding to $\Ms \simeq
3\times 10^7\msun$, marks the transition between the gas-poor dE 
galaxies and the gas-rich dI galaxies.
\item
A sharp lower bound for haloes that can form galaxies, at $V \simeq 10\kms$,
arises from the sharp drop in the cooling rate 
below $T\simeq 10^4$K, where, in the absence of metals,
it relies on molecular Hydrogen (see Barkana \& Loeb 2001, Fig.~12).  
The $H_2$ molecules are dissociated by the weak UV flux
from the first stars or AGNs long before $\zion$ 
(Haiman, Rees \& Loeb 1996;
Haiman, Abel \& Rees 2000),
allowing no gas cooling in these small dark haloes
(see DS, Figs.~5,6). We find a hint for this lower bound to $V$ in the dwarf
spheroidals of the Local Group, \Fig{V}.
\end{enumerate}

\subsection{HH galaxies}

Therefore, 
the cooling upper limit and the supernova scale limit the stellar masses 
of bright galaxies to the range $3\times 10^{10}<\Ms < 10^{12}\msun$. 
A significant fraction of the gas is assumed to have turned into stars 
in these galaxies, such that $\eta$ is not significantly correlated with 
the halo properties.
Then the tight TF relation, the high surface brightness and metallicity, 
and the weak correlation between the last two and stellar mass, all
follow naturally from the simplest possible assumptions as described in 
\se{HSB} 
(see also, e.g., Blumenthal \etal 1984). 
In summary, the assumptions made are:
\begin{enumerate}
\item
The halo is in virial equilibrium after spherical collapse 
from a cosmological background.
\item
The epoch of galaxy formation is only weakly correlated with halo mass,
consistent with the $\Lambda$CDM cosmology where the power index of density
fluctuations is $n \lsim -2$ in the range corresponding to HH galaxies.
\item
The stellar mass is proportional to the total mass, $\Ms \prop M$, 
such that 
$\eta$, the ratio of stellar to initial-gas mass, 
is uncorrelated with the halo mass. This assumption distinguishes the
HHs from the LLs. 
\item
The size of the stellar system is related to the halo virial radius 
by conservation of angular momentum, namely 
via the spin parameter $\lambda$, which is uncorrelated with the halo mass.
\end{enumerate}
The idealized picture
is disk formation via gas contraction in pre-formed haloes, 
unperturbed by recent strong galaxy interactions.
This should be especially valid for the fragile 
LL disks discussed later, 
where it is supported by a weak spatial correlation observed between 
LSBs and other galaxies, especially below a pair separation of $2\hmpc$
(Mo, McGaugh \& Bothun 1994).

It is not hard to understand why the galaxies at the bright end, mostly
large ellipticals,
are dominated by a spheroidal stellar component with little gas and a low 
current star formation rate (SFR).
The high density of the cooled gas in the early progenitors of these haloes 
allows the formation of molecules which provide efficient cooling even 
after the gas has cooled to below $10^4$K. This explains the high 
SFR in discs early on.  
Mergers of discs lead to bulges and elliptical
galaxies, which therefore tend to be those galaxies that dominate
the high $\Ms$ end.  The decrease in number of objects due to mergers
may partly explain the low scatter in radius and surface brightness 
at a given $\Ms$ in the HH regime, as indicated by K03.
The mergers provide an additional trigger for a high early SFR.
The associated high gas consumption in these early epochs results in gas-poor
systems with low SFR today.
Since the supernova feedback energy is weak compared to the depth of the
potential wells in HHs, it has a negligible effect on the SFR.

A potential caveat in the picture that assumes no gas loss from HHs 
may arise from the indications that the baryonic fraction in these systems 
may in fact be lower than the universal fraction by a factor of 2 or more 
(e.g., Klypin, Zhao \& Somerville 2002, based on semi-analytic modeling of the
Milky Way within the $\Lambda$CDM scenario, and references therein).
Such gas loss may be expected if the HHs result from a hierarchical
merger process, where the gas is lost at early stages from the small 
building blocks by supernova and radiative feedback.
Another possibility is that supernova feedback is actually stronger
than assumed, either due to microscopic effects such as porosity
in a multiphase ISM, or due to hypernova from very massive stars (Silk 2003).
Alternatively, there might be an even stronger feedback mechanism
at work in big galaxies and in clusters.
Hints from SDSS for a correlation between HHs and AGN activity 
(Kauffmann \etal 2003, in preparation), together with the established presence
of massive black holes in early-type galaxies
and the known energetic radio jets associated with AGNs,
may provide a clue for the required energetic feedback process.

\subsection{LL and dwarf galaxies} 

Most of the galaxies and most of the mass belong to the 
LL 
and dwarf family below the transition scale: $\Ms < 3\times 10^{10}\msun$.
Their halo velocities are below the critical supernova scale of 
$\Vsn \sim 100\kms$ and they are therefore subject to supernova feedback 
effects which can determine their characteristic scaling relations, 
as argued based on the simplest possible model in \se{LSB}. The energy
fed into the gas leads to a lower stellar mass fraction $\Ms/M$ 
and therefore lower surface brightness and metallicity in haloes of lower $V$. 
Some of the gas may be blown out and some may be retained or
may fall in at a later time. This gas is kept hot and possibly turbulent
such that the SFR is regulated
by supernova feedback, as well as by radiative feedback at the lower part
of the dwarf sequence.   Note that for the scaling relations to be valid
in the 
LL 
regime the gas does not have to be blown away --- it should
just be prevented from forming stars too efficiently.
Our feedback model predicts $\Ms/\Mg \prop V^2$, where $\Mg$
is the mass of the gas affected by feedback and prevented from forming stars. 
In 
LLs 
and dwarf irregulars, a significant fraction of this gas
must have been retained in the galaxy rather than been blown away.
In this case, the model indeed predicts an increasing
gas-to-star ratio for decreasing halo mass, as observed.

Our key assumption for supernova feedback is that $\Esn \prop \Ms$.
It is crudely justified also in the presence of significant radiative cooling,
when the gas is at $T \sim 10^5$K, based on the analysis of supernova remnants
by DS.
The second assumption is the straightforward energy requirement for affecting
most of the original gas, $\Esn \prop \Mg V^2$.
Together they yield that the effectiveness of feedback varies along the 
LL 
sequence as 
\be
\Ms/M \prop V^2 \, .
\label{eq:eta_v2}
\ee
The scaling relations for 
LLs 
are then obtained using
the same standard assumptions as used for HHs, namely
virial equilibrium after spherical collapse and angular-momentum conservation,
noting that the correlation of formation time with halo mass 
is even weaker for dwarfs where $n \rightarrow -3$. 
Our basic energy condition is clearly based on a simplistic model for feedback,
which was expected to provide rough estimates at best.
The fact that this model recovers so well the observed scaling relations 
is partly a matter of lucky coincidences and it should not be taken too 
literally.
However, our main moral from the remarkable success of the crude model 
is that supernova feedback can be the primary physical process 
determining the fundamental line of 
LL/dwarf 
galaxies.
One may in fact reverse the logic and infer the feedback energy relation,
$\Esn \prop \Ms \prop \Mg V^2$, from the observed scaling relations,
via the other standard assumptions of virial equilibrium and spherical
collapse in $\Lambda$CDM cosmology.
Our toy analysis therefore provides the basis and the motivation for 
detailed future studies of the supernova feedback effects, using more
sophisticated modeling and simulations. 
The inferred energy relation should serve as a useful constraint that must be 
obeyed by these models.
It may be a non-trivial challenge for such realistic models to achieve 
a match with observations as good as the match achieved by the naive
toy model.

The actual supernova feedback process is likely to be much more complex
than assumed in our toy model. For example, suprenovae exploding in a disk 
would affect the disk gas and the halo gas in different ways and
in an aspherical configuration, with fountains punching out
the ISM and the IGM in a nonuniform and possibly porous manner 
(e.g., MacLaw \& Ferrara 1999;
Scannapieco, Thacker \& Davis 2001; Mori, Ferrara \& Madau 2002; Scannapieco,
Ferrara \& Madau 2002; Silk 2003).
This would affect the way the ISM and IGM are enriched with metals
(e.g., Madau, Ferrara \& Rees 2001; Thacker, Scannapieco, \& Davis 2002).
Another complication is that the supernova energy can be transferred to 
the gas in either bulk kinetic energy or thermal energy, but it can also 
be kept in reservoirs of other forms such as turbulence, which would 
amplify the feedback effects on the gas (e.g., 
Efstathiou 2000; 
Thacker \& Couchman 2000, 2001; 
Springel 2000; Springel \& Hernquist 2003a,b).
These preliminary studies would hopefully guide us to a reliable,
realistic, detailed feedback model, but our results indicate that
the global energy balance should somehow mimic the predictions of the
early naive toy models (Larson 1974; DS).  

If supernova feedback is the primary parameter varying along
the fundamental line,
the other factors affecting the galaxy properties can be regarded
as secondary parameters, responsible for the scatter about the
mean scaling relations. These may include the following:
\begin{enumerate}
\item
The spin parameter $\lambda$ and the internal angular-momentum distribution
    for the dark matter and in particular for the baryons 
    (e.g., Dalcanton, Spergel \& Summers 1997). 
\item
The halo density profile, e.g., as parameterized by its concentration 
  parameter (Navarro, Frenk \& White 1997; see Bullock \etal 2001a).
\item
The epoch when the halo collapsed relative to the typical collapse time
  of haloes of a similar mass and its mass accretion/merger history.
\item
The detailed efficiency of gas cooling and gas removal by feedback.
\item
The star formation efficiency, e.g., as a function of gas surface density. 
\end{enumerate}
In general, the surface brightness at a given $\Ms$
is expected to be below the mean relation for younger galaxies that had
more recent major mergers and therefore typically have high spin 
and low concentration (e.g. Wechsler \etal 2002).
These qualitative predictions are consistent with the findings from
the SDSS data as well as the Local-Group dwarfs.

As mentioned in \se{general}, the distribution of spin parameter, 
in particular, is a natural cause for variation
in stellar radius, and therefore surface brightness of disk galaxies
(see modeling by Fall \& Efstathiou 1980; Dalcanton, Spergel \& Summers 1997;
Mo, Mao \& White 1998; van den Bosch 2000, 2001;
Avila-Reese \& Firmani 2000, 2003; Firmani C. \& Avila-Reese 2000).
All the scatter observed in $\Rs$ and $\mus$ for SDSS galaxies (K03) 
can be accounted for by scatter in halo spin parameter
(e.g., Bullock \etal 2001b).
However, the lack of correlation between $\lambda$ and halo mass
implies that the spin parameter cannot be the primary factor driving
the systematic variation of $\mus$ with halo mass for 
LL/dwarfs. 

It is worth noting though that the key relation leading to 
the observed scaling relations, $\Ms/M \prop V^2$, might in principle arise 
from processes other than feedback. For example, $\Ms/M$ may increase with 
halo mass because of a systematic increase in star formation efficiency due 
to the higher gas surface density (e.g., Galaz \etal 2002). 
Is this enough for explaining the observed correlation without
a contribution from feedback effects, namely such that the cold-gas mass
is $\prop M$?
Consider a star formation rate which depends on gas surface density as in
the Kennicutt-Schmidt law, $\dot \mus \prop \mug^N$, namely
$\dot\Ms/M \prop \mug^{N-1}$.
From the virial relation we have from \equ{mus_n} and \equ{vir_n}
$\mug \prop \lambda^{-2} M^{\tau} \propto \lambda^{-2} V^{3\tau}$
with $0\leq \tau \leq 1/3$ (for $-2 \geq n \geq -3$).
We therefore obtain $\dot\Ms/M \prop V^{3\tau(N-1)}$.
If we assume $\dot\Ms \prop \Ms$, as indicated for the SDSS
LLs 
and the Local-Group dIs, we get the required scaling
relation $\Ms/M \prop V^2$ for $3\tau(N-1)=2$. With $\tau \leq 1/3$
we obtain $N\geq 3$. This is a stronger dependence of SFR on surface
gas density than measured for star-forming galaxies, $N=1.4\pm 0.15$ 
(Kennicutt 1998). It indicates that star-formation efficiency is not 
a natural primary driver for the 
LL 
scaling relations, though it may have an important role.

\subsection{Related work on the role of feedback}

Our model for the additional role of radiative feedback in distinguishing
dwarf spheroidals from dwarf irregulars can be regarded as a qualitative 
speculation, to be investigated in more detail in future work.

The trend of increasing total-mass-to-light ratio with decreasing mass,
which naturally results from supernova feedback and reproduces the fundamental
line, may not be enough for fully resolving the mystery of missing dwarf 
galaxies (Klypin \etal 1999; Moore \etal 1999; Springel \etal 2001). 
The discrepancy is not only between the faint-end slopes
of the predicted mass function and the observed luminosity function,
but it also involves the distribution of internal velocities. 
The halo masses inferred straightforwardly from the observed velocities 
in the LG dwarf galaxies, many of which are on the order of $\sim 10\kms$,
are too small compared to the CDM model predictions.
One possibility is that the relevant halo velocities are severely 
underestimated because the sampling by stars in dwarfs is biased towards
the very inner halo regions, where the rotation and dispersion velocities 
may be significantly lower than the maximum or virial velocity relevant for
mass estimation (Stoehr \etal 2002). This possibility is unlikely to
provide the full answer because the velocities measured from HI gas, which
typically samples more extended radii by a factor of 2-3, are still 
of similarly low, $\sim 10\kms$ (Blitz 2003, private communication).
The discrepancy between the observed velocity function and that predicted
by the CDM model thus seems to indicate the presence of some barren haloes, 
which are completely dark and show no trace of luminous stars in them
(e.g., Kochanek 2001).
Attempts have been made to explain the barren haloes
by the radiative feedback effects discussed in \se{dE},
which indeed are expected to ``squelch" the formation of stars
in haloes that form after cosmological reionization at
$z\sim 7$ (e.g., Bullock, Kravtsov \& Weinberg 2000; Somerville 2002;  
Tully \etal 2002). 
Such barren dark haloes may alternatively be explained by the destructive 
effect of energetic outflows from one galaxy on neighboring forming
protogalaxies via ram pressure brushing aside the tenuously held gas
(e.g., Scannapieco, Ferrara \& Broadhurst 2000; Scannapieco \& Broadhurst
2001; Scannapieco, Thacker \& Davis 2001; Thacker, Scannapieco \& Davis 2002).
These processes are yet to be studied with more realistic simulations,
in an attempt to find out whether they can indeed explain the absence of 
luminous components in massive enough haloes.


A scale similar to the supernova scale, originating in a different way 
from the features of the cooling curve,
is associated with another transition between different behaviors,
as discussed in Birnboim \& Dekel (2003).
They found, using analytic arguments supported by simulations with
a 
spherically 
symmetric Lagrangian hydrodynamical code,
that in haloes less massive than $\sim 3\times 10^{11}\msun$
the gas falling into the halo does not cross a virial shock until 
it hits the ``disc"
itself. The ``standard" virial shock develops only in more massive haloes, 
hosting large galaxies and clusters, where the shock quickly
expands to near the virial radius. Then, as commonly assumed,
infalling gas is heated behind
the shock to the halo virial temperature
and is kept pressure supported in the halo until it cools radiatively
and slowly contracts into the disc. In less massive haloes, where the virial
temperature is below a few $\times 10^5$K, the shock that tries to develop
loses energy very efficiently via radiation that is dominated by He
recombination and Oxygen lines.
This prevents the shock from ever expanding into the halo.
A possible implication of this result is that early star formation
becomes more efficient in haloes of $M<3\times 10^{11}\msun$,
in shocks produced by the cold infalling gas when it hits the cold gaseous
disc, giving rise to the burst which heats much of the remaining gas and
produces an LSB galaxy. Further infalling gas may prevent blowout
and keep the hot gas in the galaxy, giving rise to gas-rich dwarf
irregulars.

The low baryonic fraction observed in $V \sim 60\kms$ LSB's (by BBS),
together with the prediction of $\eta \prop V^2$ for the feedback effect,
implies $\Vsn \sim 80\kms$, which is quite consistent with the DS estimate
of $\Vsn \sim100\kms$ and with the observed transition scale at a stellar
mass $\Ms \gsim 10^{10}\msun$.
Furthermore, as argued by Maller \& Dekel (2002), feedback can help solve
the apparent angular-momentum problem within the CDM scenario, where the 
baryons in cosmological simulations seem to lose most of their angular momentum
and fail to form large discs as observed (Navarro \& Steinmetz 2000).
Maller, Dekel \& Somerville (2001) and
Maller \& Dekel (2002) modeled the properties of the
LSB galaxies observed by BBS based on spin buildup from the orbital angular
momenta along the halo merger history, combined with gas blowout from small 
merging satellites and the associated baryonic spin increase. They
found that a value of $\Vsn \sim 90\kms$ can indeed explain the
higher spin observed in the LSB galaxies. 

We note that gas blowout in small haloes may also help resolve a third
problem of the CDM scenario, the 
cusp/core problem of halo density profiles, where simulated haloes show a 
steep inner cusp while observations indicate that at least some galaxies 
have flat-density cores. While feedback cannot significantly
affect the dark-matter distribution in big galaxies,
an impulsive blowout may reduce the core densities in dwarf satellites
by a factor of a few (Gnedin \& Zhao 2002).  
When these puffed-up satellites merge to build up bigger haloes,
they get tidally disrupted before they manage to penetrate the inner regions
and turn the cores into cusps. In this indirect way the feedback can
help the survival of cores even in relatively big galaxies 
(Dekel, Devor \& Hetzroni 2003; Dekel \etal 2003). 
However, working out a feedback model within the CDM scenario that will 
explain the possible existence of cores in giant galaxies and clusters 
of galaxies could be challenging; it will require a feedback mechanism
more energetic than simple supernova feedback, perhaps by hypernovae from
massive stars or by radio jets from AGNs.

We conclude that
feedback effects seem to be able to provide the cure to all three 
major problems facing galaxy formation theory within the CDM scenario.
Understanding the details of the feedback processes is therefore
a major goal of galaxy formation studies.

\section*{Acknowledgments}
We acknowledge stimulating discussions with S. Courteau,
G. Kauffmann, C. Kochanek,
J.P. Ostriker, D. Spergel, N. Shaviv \& S.D.M. White.
This research has been supported
by the Israel Science Foundation grant 213/02,
by the US-Israel Bi-National Science Foundation grant 98-00217, 
by the German-Israel Science Foundation grant I-629-62.14/1999,
and by NASA ATP grant NAG5-8218.


\label{lastpage}
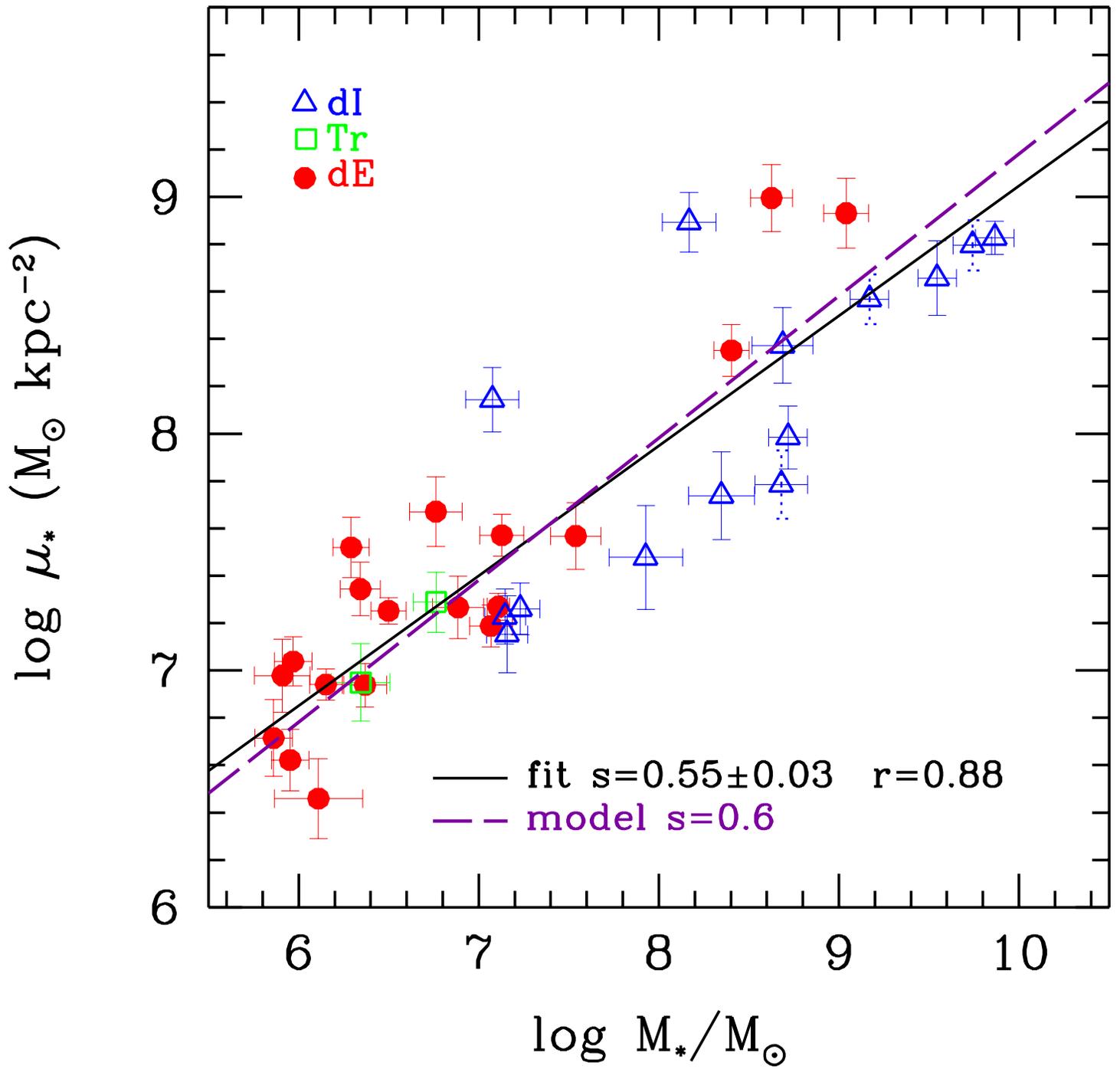
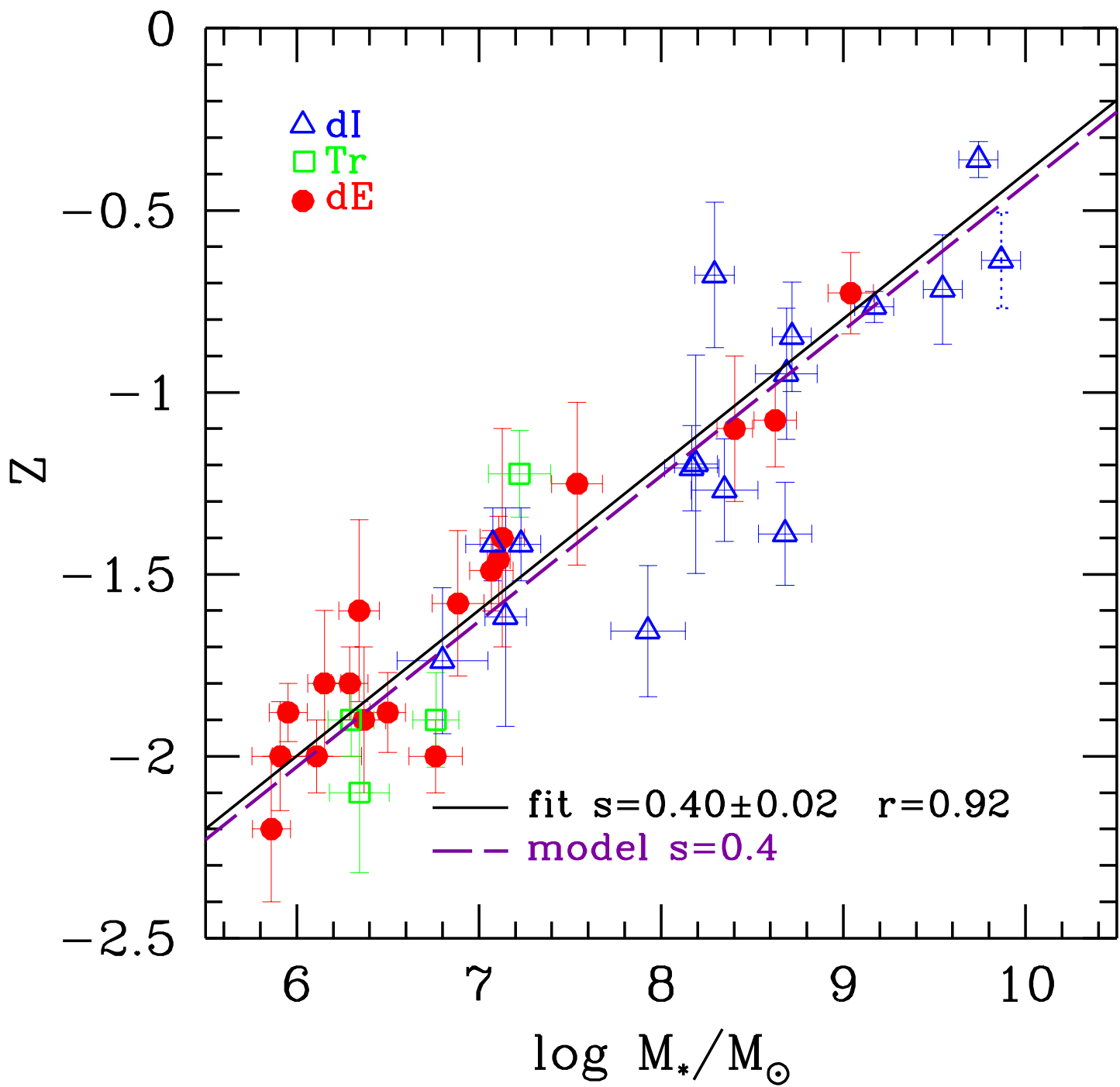
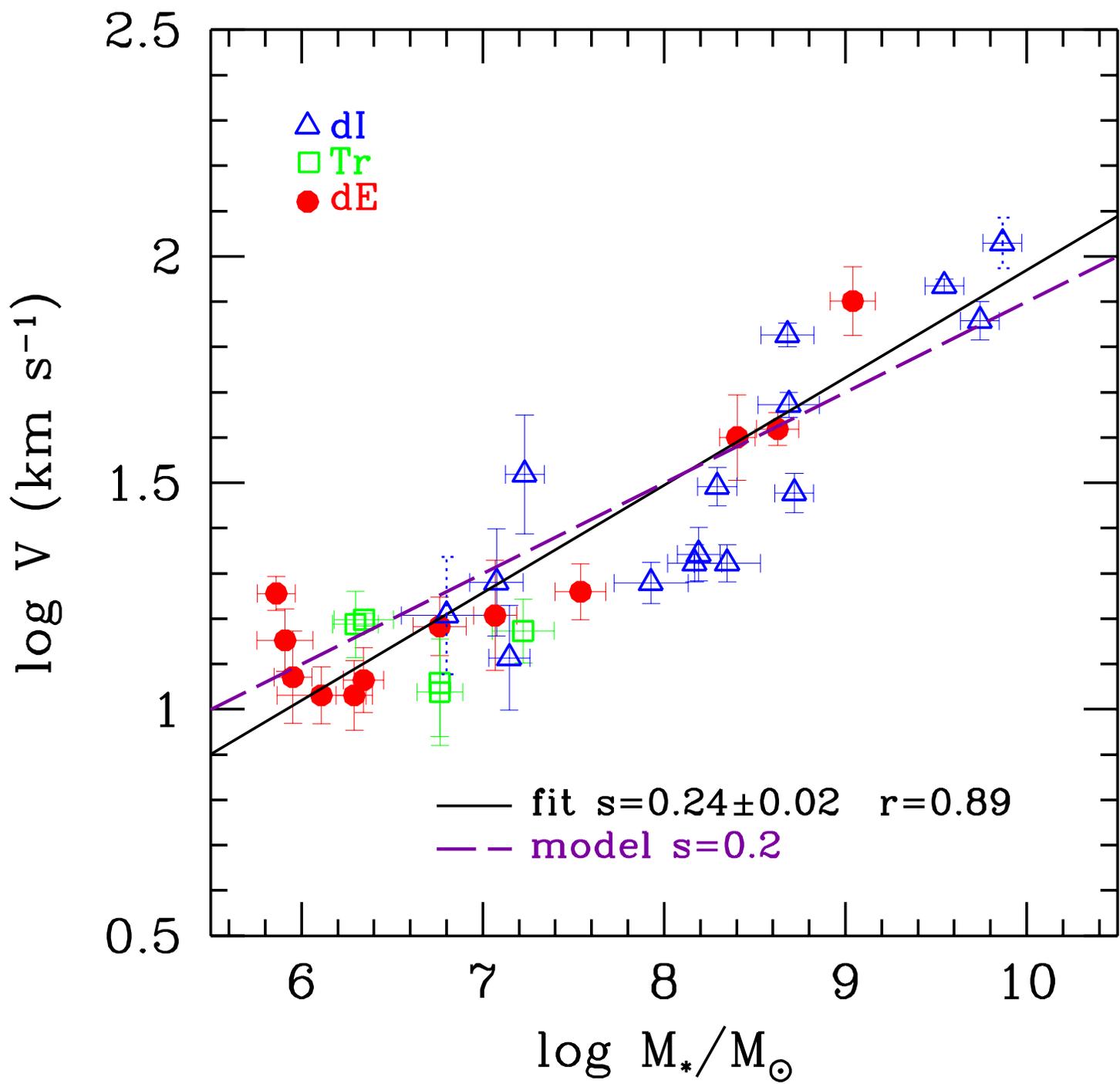
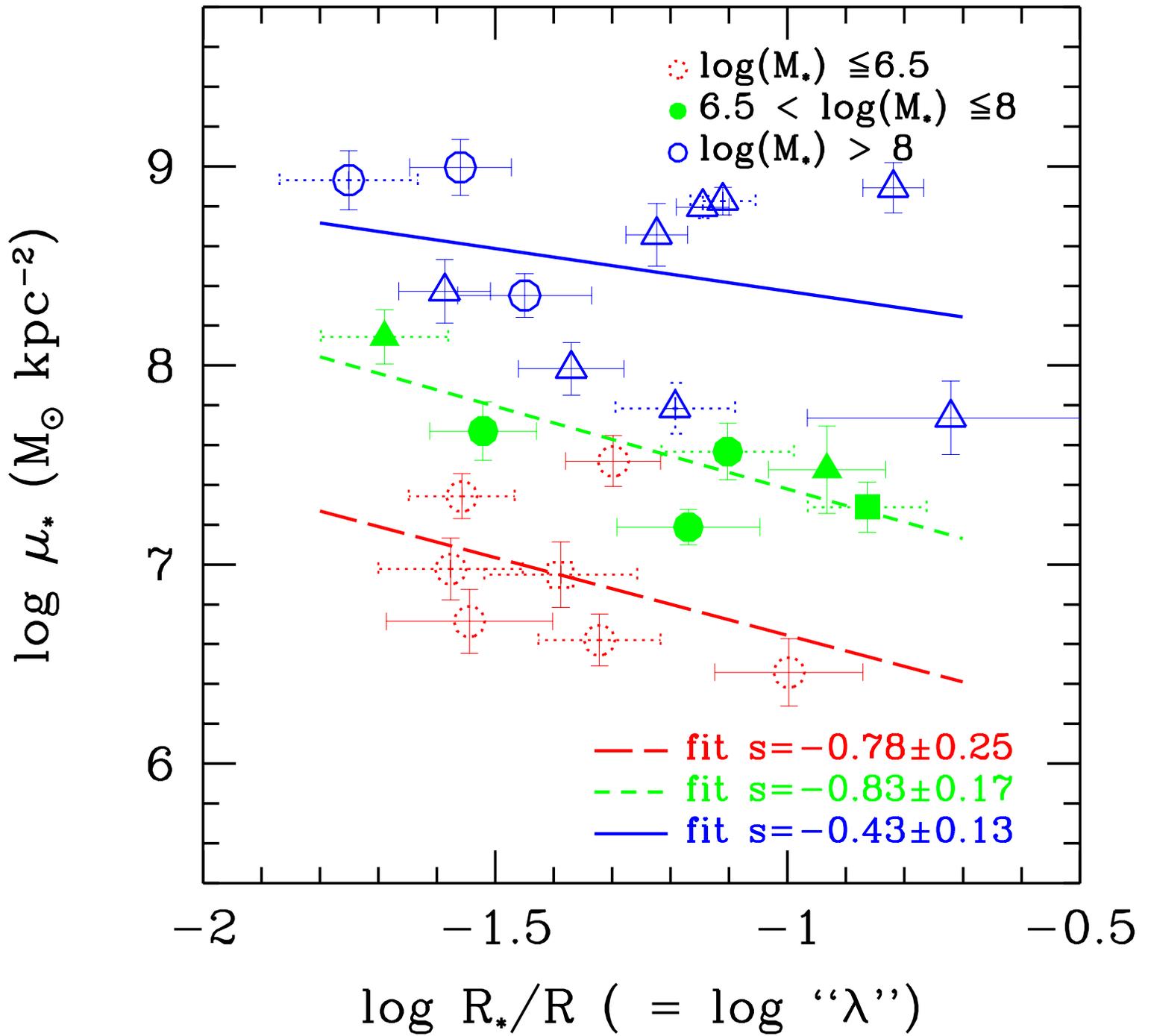
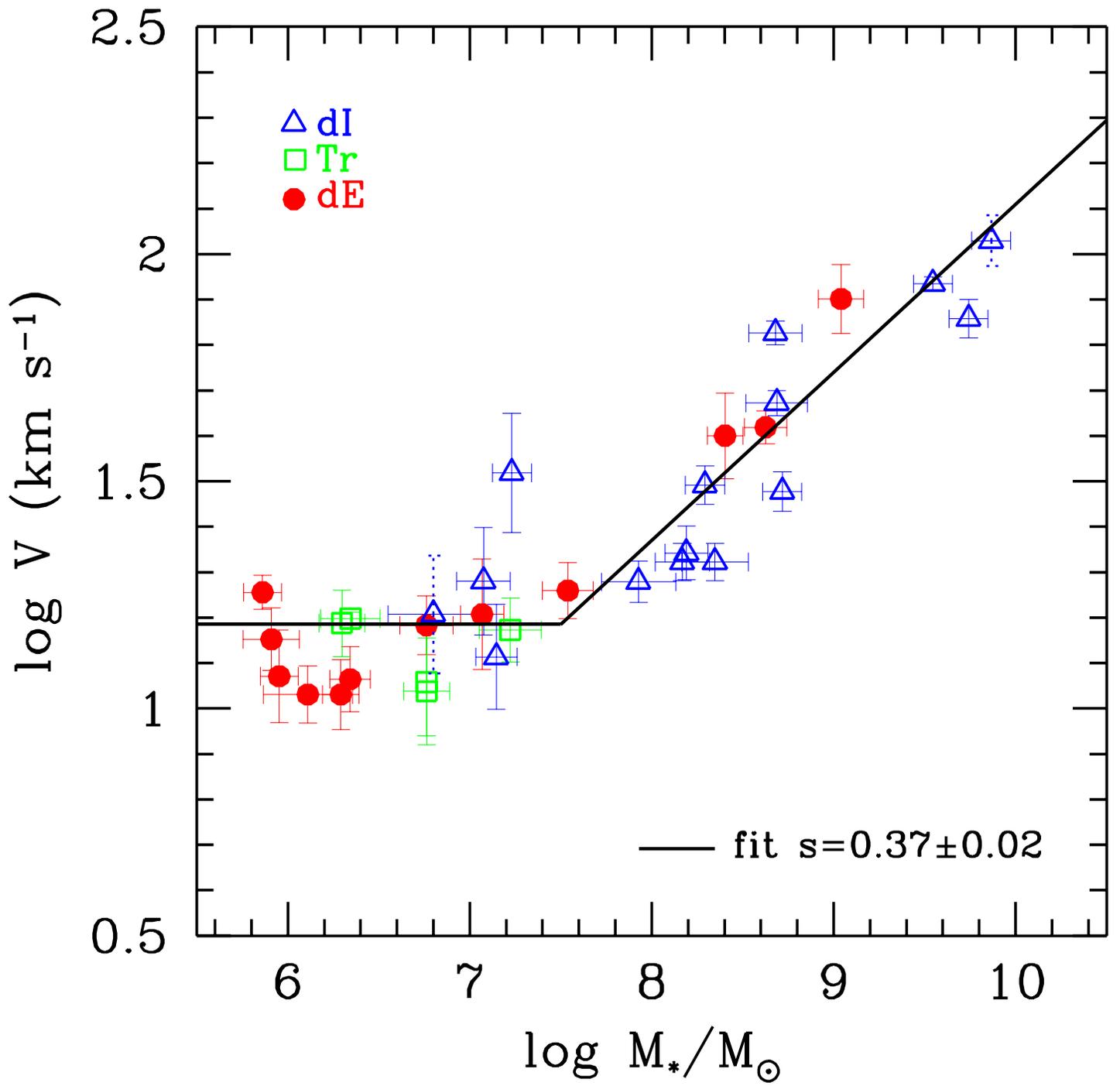
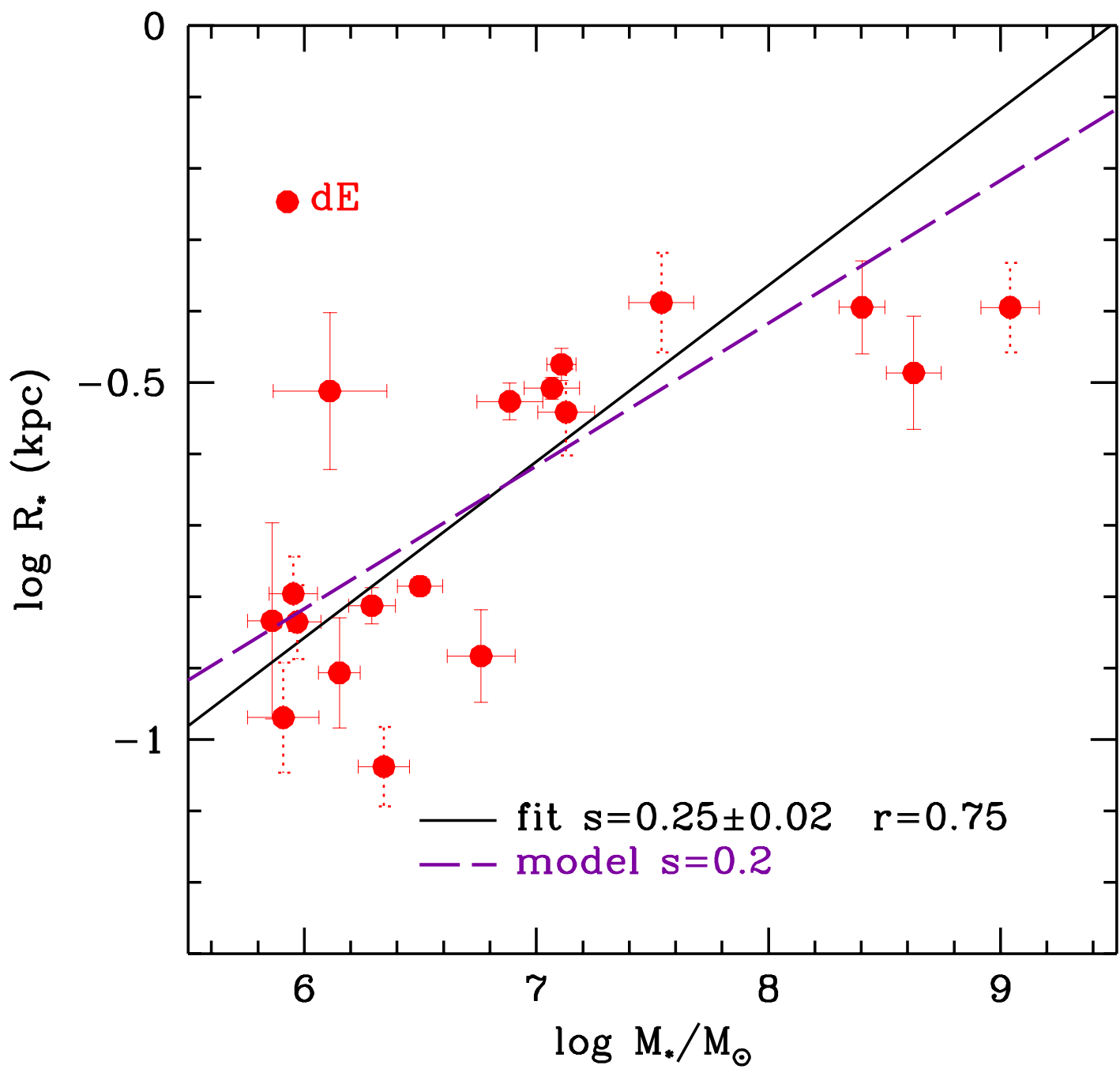
\end{document}

Bernardi et al 0110344 V FP 
Tremanti (Heckman) Z  

astro-ph/0101458 Schombert, McGaugh, Eder  gas/mass dwarfs
Garnett (astro-ph/0209012)   V-L-Z in S and dI
Heckman 00, Pettini 01   wind = sfr

Salucci 

Governato astro-ph/0207044 J-catastrophe etc.

Blanton  0209479
correlations 
Eisenstein spectra

-Bothun, Impey \& McGaugh (1997): review.
 Fig 4.  shows spread in mu, but <mu> decreasing with dec L
 Fig 1.  SB function flat. But can be due to low L.  consistent with mu-L cor.
 Fig 3.: W_50 dist.
 LSB large and luminous:
 Malin 1 type. But angular-diameter limited surveys are strongly biased towards
 finding the largest galaxies.
-O'Neil \& Bothun 2000   SB function flat in the range 22-25.

-Impy \etal (1996) catalog
-O'Neil K., Bothun G., Schombert J., 1999, catalog

-Phillipps, S.; Davies, J. I.; Disney, M. J.  1988MNRAS.233..485P:
Is there really a luminosity-sb relation for dwarf galaxies?
	        
-Dalcanton, Spergel \& Summers (1997): LSB are in large part a continuous
extention of the population of normal disk galaxies, reaching
to higher angular momenta and lower masses. 
(we focus here on lower masses.  lambda secondary).

-LSB are more gas rich than normal late spirals: Longmore \etal (1982)
and bluer: 
McGaugh 1992; de Blok, van der Hulst \& Bothun (1995)
Sprayberry \etal (1995)
-low gas metallicity: McGaugh (1994) de Blok \& McGaugh (1997)

-extended HI disks, low gas surface density and high M/L:
van der Hulst \etal (1993)
de Blok, McGaugh \& van der Hulst (1996)
-Little effect by external influences on speeding up the evolution
McGaugh \& de Blok (1997)
la